\documentclass[11pt,letterpaper]{article}
\usepackage[latin1]{inputenc}
\usepackage{amsmath}
\usepackage{amsfonts}
\usepackage{amssymb}
\usepackage{graphicx}
\usepackage{setspace}
\usepackage{graphicx}
\usepackage{bm}
\usepackage{epstopdf}
\usepackage{mathrsfs}
\usepackage{bbding}
\usepackage[tight,TABTOPCAP]{subfigure}
\usepackage{pict2e}
\usepackage{float}
\usepackage{undertilde}
\usepackage[english]{babel}
\usepackage[abs]{overpic}
\usepackage[left=2cm, right=2cm, top=2.50cm, bottom=2.00cm]{geometry}

\newcommand{\cone}{$\mathscr{C}^1$}

\renewcommand{\newline}{\medskip}

\newcommand{\der}{\ensuremath{\rm d}}
\newcommand{\vect}[1]{\pmb{#1}}

\newcommand{\grad}{\ensuremath{\nabla}}

\newcommand{\mOne}{M$_1$ }
\newcommand{\mTwo}{M$_2$ }
\newcommand{\mThree}{M$_3$ }
\newcommand{\pooz}{[110] }

\pdfoutput=1

\begin{document}

\begin{center}
\Large{Shape Memory Alloy Nanostructures With \\Coupled Dynamic Thermo-Mechanical Effects} 
\end{center}

\begin{center}
R. Dhote$^{1,3}$, H. Gomez$^2$, R. Melnik$^3$, J. Zu$^1$

$^1$Mechanical and Industrial Engineering, University 
of Toronto, \\5 King's College Road, Toronto, ON, M5S3G8, Canada\\
$^2$Department of Applied Mathematics, University of A Coru\~{n}a \\ Campus de Elvina, s/n. 15192 A Coru\~{n}a, Spain\\
$^3$ The MS2Discovery Interdisciplinary Research Institute, \\
M$^2$NeT Laboratory, Wilfrid Laurier University, Waterloo, ON,  
N2L3C5, Canada
\end{center}

\textit{Abstract:}
Employing the Ginzburg-Landau phase-field theory, a new coupled dynamic  thermo-mechanical 3D model has been proposed for modeling the cubic-to-tetragonal martensitic transformations in shape memory alloy (SMA) nanostructures. The stress-induced phase transformations and thermo-mechanical behavior of nanostructured SMAs have been investigated. The mechanical and thermal hysteresis phenomena, local non-uniform phase transformations and corresponding non-uniform temperature and deformations distributions are captured successfully using the developed model. The predicted microstructure evolution qualitatively matches with the experimental observations. The developed coupled dynamic  model has provided a better understanding of underlying martensitic transformation mechanisms in SMAs, as well as their effect on the thermo-mechanical behavior of nanostructures. \\
\section{Introduction}
Shape memory alloys (SMAs) exhibit a remarkable hysteresis behavior that arises from an interplay between microstructures in two phases, a high temperature higher symmetric austenite and low temperature lower symmetric martensite phases. The phase transformations between two phases, called martensitic transformations (MTs), reveal non-uniform and temperature dependent microstructure morphology. The microstructure morphology manifests itself through a domain pattern \cite{bhattacharya2003microstructure} and surface relief \cite{wu2009reversible} due to elastic deformations (or strains) under temperature or stress induced loadings. The ability to control deformations in different geometries and loading conditions can lead to tailoring of mechanical structures \cite{bhattacharya2005material} for better SMA-based actuators and sensors in NEMS, MEMS and biomedical applications \cite{vzuvzek2012electrochemical,koig2010micro,bayer2011carbon,kohl2004shape,miyazaki2009thin,yoneyama2009shape}. The ability to control deformations in SMAs as well as in other functional materials and semiconductor devices \cite{prabhakar2010influence,scott2007applications} can be better understood by physics-based modeling approaches. The physics-based models facilitate the study of 3D complex martensitic twin microstructures and properties of SMAs.

Recently, MTs have been studied by using atomistic simulations \cite{zhong2011atomistic}. Simulating larger domains at submicron scales requires enormous computational power which limits their use. In order to overcome this limitation, continuum descriptions at the submicron length scale have become an effective tool to model experimentally observed phenomena. One of the continuum based methods is the phase-field (PF) approach that has been widely used to investigate qualitatively microstructures and underlying mechanisms in nano-ferroic systems \cite{Bouville2008,ng2012domain}. Particularly, in the ferroelastic systems, the PF model has been implemented successfully to understand martensitic transformations in SMA nanostructures under isothermal \cite{Ahluwalia2006} or athermal conditions \cite{levitas2007athermal}. The isothermal assumption particularly holds for quasistatic or slow strain rate loadings.  However, the isothermal assumption is a strong hypothesis under a dynamic loading of SMAs as the phase transformation causes the temperature evolution due to self-heating or cooling of the system, consequently  affecting their mechanical behavior, due to insufficient time for heat transfer to dissipate in the environment as observed experimentally \cite{shaw1997nucleation,pieczyska2006phase}. Therefore, a theoretical framework that couples dynamically the mechanical behavior with temperature evolution is imperative to describe the MTs and their behavior in SMA nanostructures for critical application developments. 

\section{Mathematical Model} 
For the first time, we develop and apply a  fully coupled dynamic thermo-mechanical 3D PF model to the investigation of the stress-induced phase transformations and behavior in SMA nanostructures. The cubic-to-tetragonal PT is described by deviatoric strains defined in terms of symmetry adapted combinations of the components of the strain tensor. The model is developed based on the Ginzburg-Landau theory. We call $\vect\epsilon$ the Cauchy-Lagrange infinitesimal strain tensor that can be defined in components as  
$\epsilon_{ij} = \left( u_{i,j} +  u_{j,i} \right)/2$, where the $u_i$'s represent the displacements and an inferior comma denotes partial differentiation. We will make use of the following strain measures for the cubic-to-tetragonal phase transformation; 
$ e_1 = \displaystyle 1/\sqrt{3} (\epsilon_{11} + \epsilon_{22} + 
\epsilon_{33} ) $, 
$ e_2 =  1/\sqrt{2} (\epsilon_{11} - \epsilon_{22} ) $,  
$ e_3 =  1/\sqrt{6} (\epsilon_{11} + \epsilon_{22} - 2 \epsilon_{33}) $,
$ e_4 = \epsilon_{23} $,  
$ e_5 = \epsilon_{13} $,  
$ e_6 = \epsilon_{12} $.
Here, $ e_1 $ is the hydrostatic strain, $ e_2,  e_3 $ are the deviatoric strains, and $ e_4 $, $ e_5 $ and $ e_6 $ are the shear strains. We use the symmetric strain-based free energy functional, for cubic-to-tetragonal phase transformations, initially proposed by Barsch et al. \cite{Barsch1984} and later modified by Ahluwalia et al. \cite{Ahluwalia2006} to study the martensitic transformations in SMA nanostructures. The free energy functional $ \mathscr{F} $ with anharmonic components of OPs and harmonic components of non-OP components is expressed as 
\begin{equation}
\mathscr{F}[\vect u] = 
\int_{\Omega} \left[F_0(e_i,\tau) +\frac{k_g}{2} \left( |\grad e_2|^2 + |\grad e_3|^2 \right) \right] \der\Omega,
\label{eq:FEcub2tet}
\end{equation}
where $|\cdot|$ denotes the norm of a vector, and $F_0$ is the function
\begin{eqnarray}
\label{homog_free}
F_0(e_i,\tau)&= \displaystyle\frac{a_1}{2} e_1^2 + \frac{a_2}{2} \left(e_4^2 + e_5^2 +e_6^2\right) + a_3 \tau  \left(e_2^2 + e_3^2\right) \nonumber \\
& + a_4 e_3 \left(e_3^2 - 3 e_2^2\right)  + a_5 (e_2^2 + e_3^2)^2.
\end{eqnarray}
Here, the $ a_i $'s and $ k_g $ are material parameters, $ \tau $ is the dimensionless temperature coefficient defined as $ \tau = \displaystyle (\theta - \theta_m)/(\theta_0 - \theta_m) $, where $ \theta_0 $ and $ \theta_m $ are the material properties specifying the transformation start and end temperatures. The kinetic energy $\mathscr{K}$, the energy associated to body forces $\mathscr{B}$, and the Raleigh dissipation $\mathscr{R}$ are defined as $\mathscr{K}[\dot{\vect u}]=\int_{\Omega}\frac{\rho}{2}|\dot{\vect u}|^2 \der\Omega,\,\mathscr{B}[\vect u]=-\int_{\Omega}\! \vect f\cdot\vect u \der\Omega,\,\mathscr{R}[\dot{\vect u}]=\int_{\Omega}\frac{\eta}{2} |\dot{\vect e}|^2\der\Omega$,

where a dot over a function denotes partial differentiation with respect to time, $\rho$ is the density, $ \vect f $ is the body load vector, $ \eta $ is the dissipation coefficient, and $\vect e$ is a vector collecting the $e_i$'s. The potential energy of the system $\mathscr{U}$ is defined as $\mathscr{U}[{\vect u}]=\mathscr{F}[{\vect u}]+\mathscr{B}[{\vect u}]$, while the Lagrangian takes on the form $\mathscr{L}[\vect u,\dot{\vect u}]=\mathscr{K}[\dot{\vect u}]-\mathscr{U}[{\vect u}]$. We define the Hamiltonian as $\mathscr{H}[\vect u,\dot{\vect u}]=\int_0^t\mathscr{L}[\vect u,\dot{\vect u}]\der {\it t}$.
Following a variational approach, the governing equation of motion has the form
\begin{equation}
\label{Ch14eqs_motion}
\frac{\partial}{\partial t}\left(\frac{\delta\mathscr{L}}{\delta\dot{u_i}}\right)-\frac{\delta\mathscr{L}}{\delta u_i}=-\frac{\delta\mathscr{R}}{\delta \dot{u_i}},
\end{equation}
where the operator $\delta(\cdot)/\delta (\cdot)$ denotes a variational derivative. Defining 
\begin{equation}
\sigma_{ij}=\frac{\partial\mathscr{F}}{\partial u_{i,j}},\, \mu_{ij,kk}=-\left(\frac{\partial\mathscr{F}}{\partial u_{i,jk}}\right)_{,k},\,\sigma'_{ij}=\frac{1}{\eta}\frac{\partial\mathscr{R}}{\partial u_{i,j}},
\end{equation}
Eq. \eqref{Ch14eqs_motion} may be rewritten as 
\begin{equation}
\label{strucural_eqns}
\rho \ddot{u}_{i} = \sigma_{ij,j} + \eta\sigma^{\prime}_{ij,j} + \mu_{ij,kkj} + f_i.
\end{equation}
Finally, the internal energy is $\iota=\Psi(\theta,\vect u)-\theta\frac{\partial\Psi(\theta,\vect u)}{\partial\theta}$ with $\Psi(\theta,\vect u)=\mathscr{L}[\vect u,\dot{\vect u}]-C_v\theta\ln(\theta)$. Stating an energy balance equation, and using Fourier's law, the following equation for the temperature evolution may be derived as
\begin{equation}
\label{thermal_eqns}
C_v\dot{\theta} =\kappa\theta_{,ii}+\Xi\theta\left(u_{i,i}\dot{u}_{j,j} -3u_{i,i}\dot{u}_{i,i}\right) + g.
\end{equation}
Here, $C_v$, $\kappa$, and $\Xi$ are positive constants that represent, respectively, specific heat, thermal conductance coefficient, and strength of the thermo-mechanical coupling, while $g$ is a thermal load. The simulations were performed for homogeneous single crystal FePd rectangular prism nanowire of dimension 160$\times$40$\times$40 nm. The material properties used during the simulations are \cite{Bouville2008, 
Kartha1995}: $a_1 $= 192.3 GPa, $a_2 $= 280 GPa,$a_3 $= 19.7 GPa, $a_4 $= 2.59$ \times $ 10$ ^3 $ GPa, $a_5 $= 8.52$ \times $ 10$ ^4$ GPa, $ k_g $ = 3.5 $ \times $ 10$ ^{-8} $ N, $\rm{\theta_m = 270}$ K, $\rm{\theta_0 = 295}$ K, ${C_v = 350}$ $\rm{J kg^{-1} K^{-1}}$, $\rm{\kappa = 78}$ $\rm{W m^{-1} K^{-1} }$, and $ \rho = 10000 $ kg m$^{-3}$. 

The governing Eqs. (\ref{strucural_eqns})--(\ref{thermal_eqns}) are strongly thermo-mechanically coupled in a  non-linear manner, with the fourth-order differential terms. These complex equations are not amenable to a closed form solution.  They pose great challenges to numerical approaches. We have developed an isogeometric analysis (IGA) \cite{Hughes} framework that allows the straightforward solution to the developed model. It also allows the use of coarser meshes, larger time steps along with geometrical flexibility and accuracy \cite{dhote2014CMAME}.

\section{Numerical Simulations} 
To elucidate the capabilities of the developed model from physics point of view, the simulations have been conducted on a rectangular prism nanowire of dimension 160$ \times $40$ \times$40 nm (meshed with 162$ \times $42$ \times$42 uniform quadratic $ \mathscr{C}^1 $-continuous B-spline basis functions) to investigate its thermo-mechanical behavior subjected to dynamic stress-induced loadings. Simulations to examine the pseudoelastic (PE) and shape memory effect (SME) behavior of SMAs, as a function of microstructures, have been performed. 

\begin{figure}[h!]
\centering
\subfigure 
{
\begin{overpic}[trim=0mm 10mm 0mm 0mm,clip, width=0.3\linewidth] {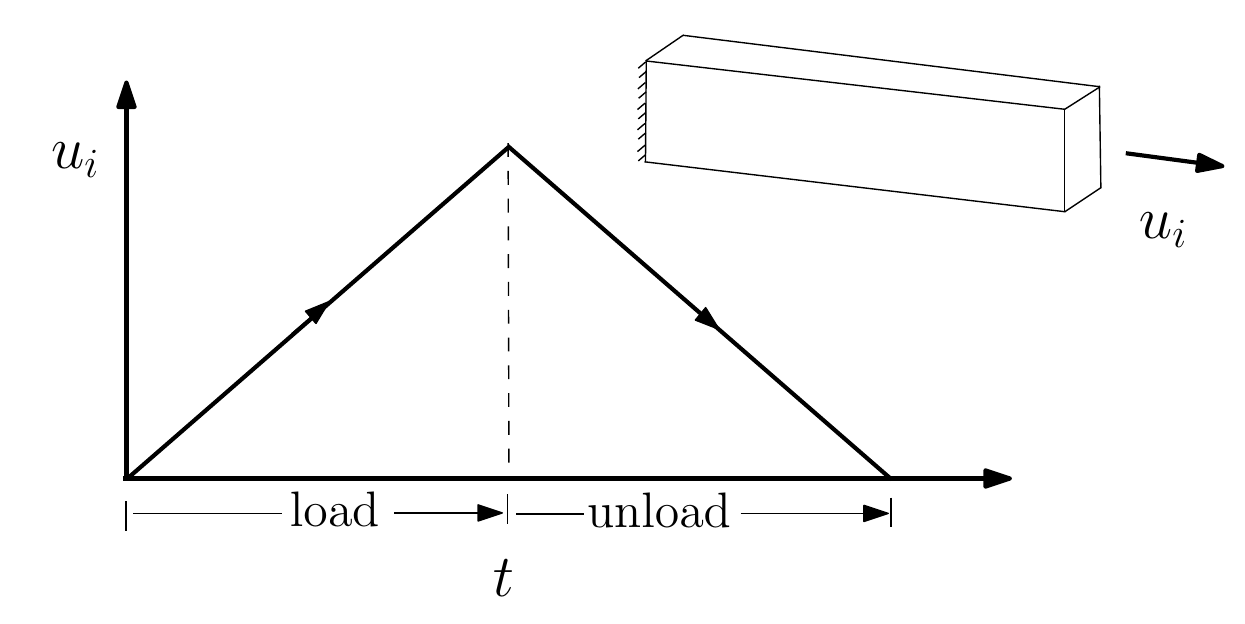}
\put(0,65){(o)}
\end{overpic}
}
\subfigure 
{
\begin{overpic}[trim=0mm 10mm 0mm 10mm,clip, width=0.3\linewidth] {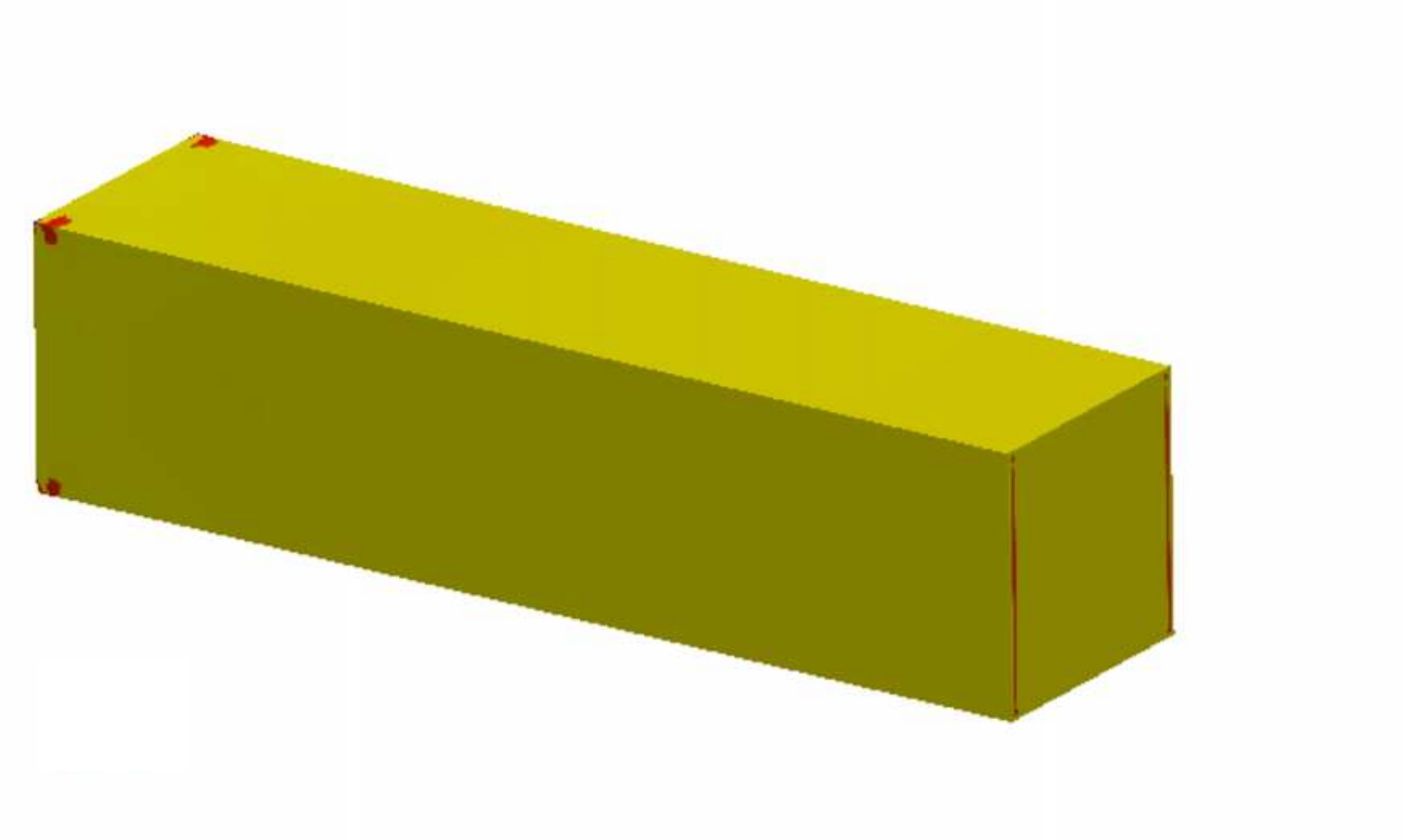}
\put(0,65){(a)}
\end{overpic}
}\\
\subfigure 
{
\begin{overpic}[trim=0mm 10mm 0mm 10mm,clip, width=0.3\linewidth] {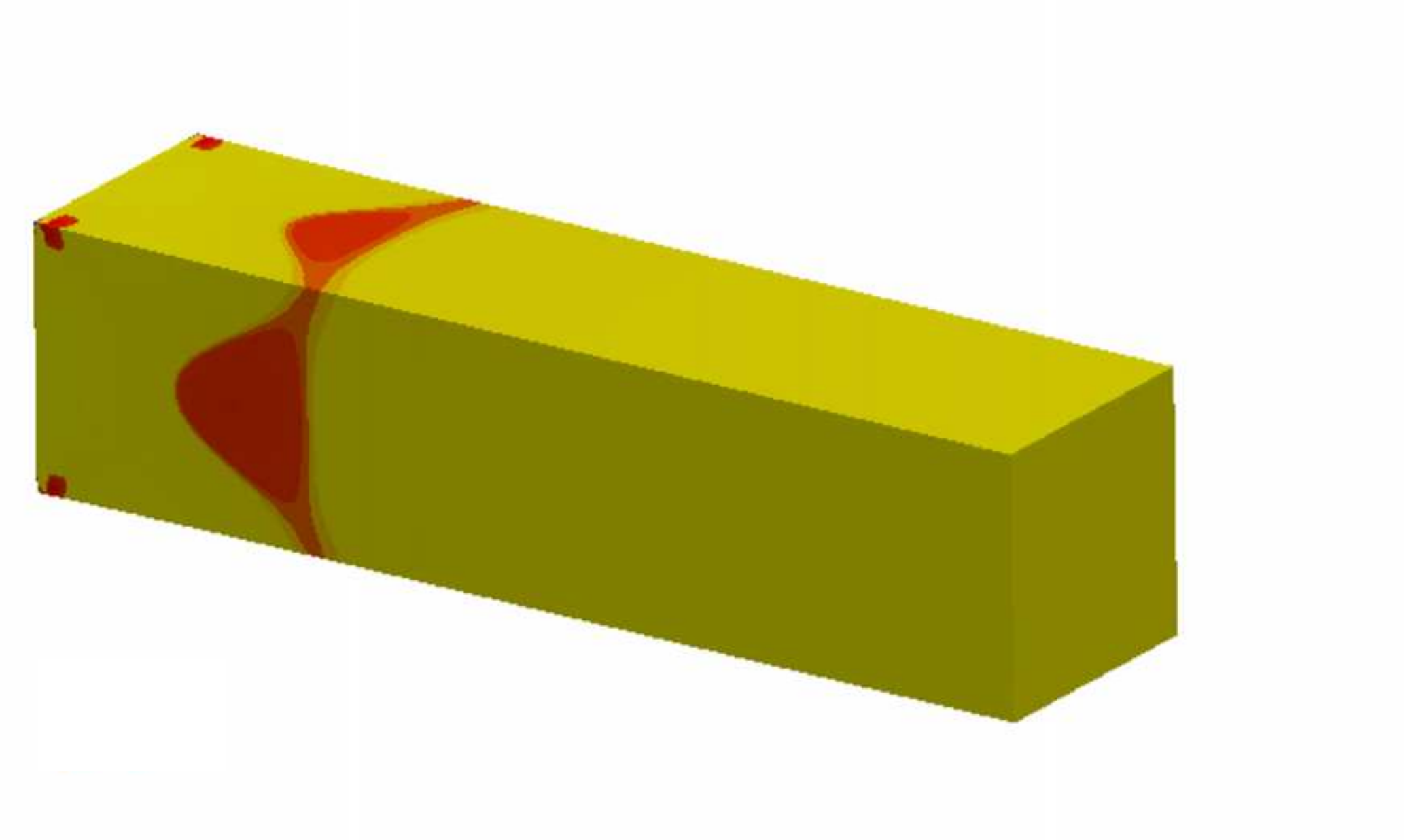}
\put(0,65){(b)}
\end{overpic}
}
\subfigure 
{
\begin{overpic}[trim=0mm 10mm 0mm 10mm,clip, width=0.3\linewidth] {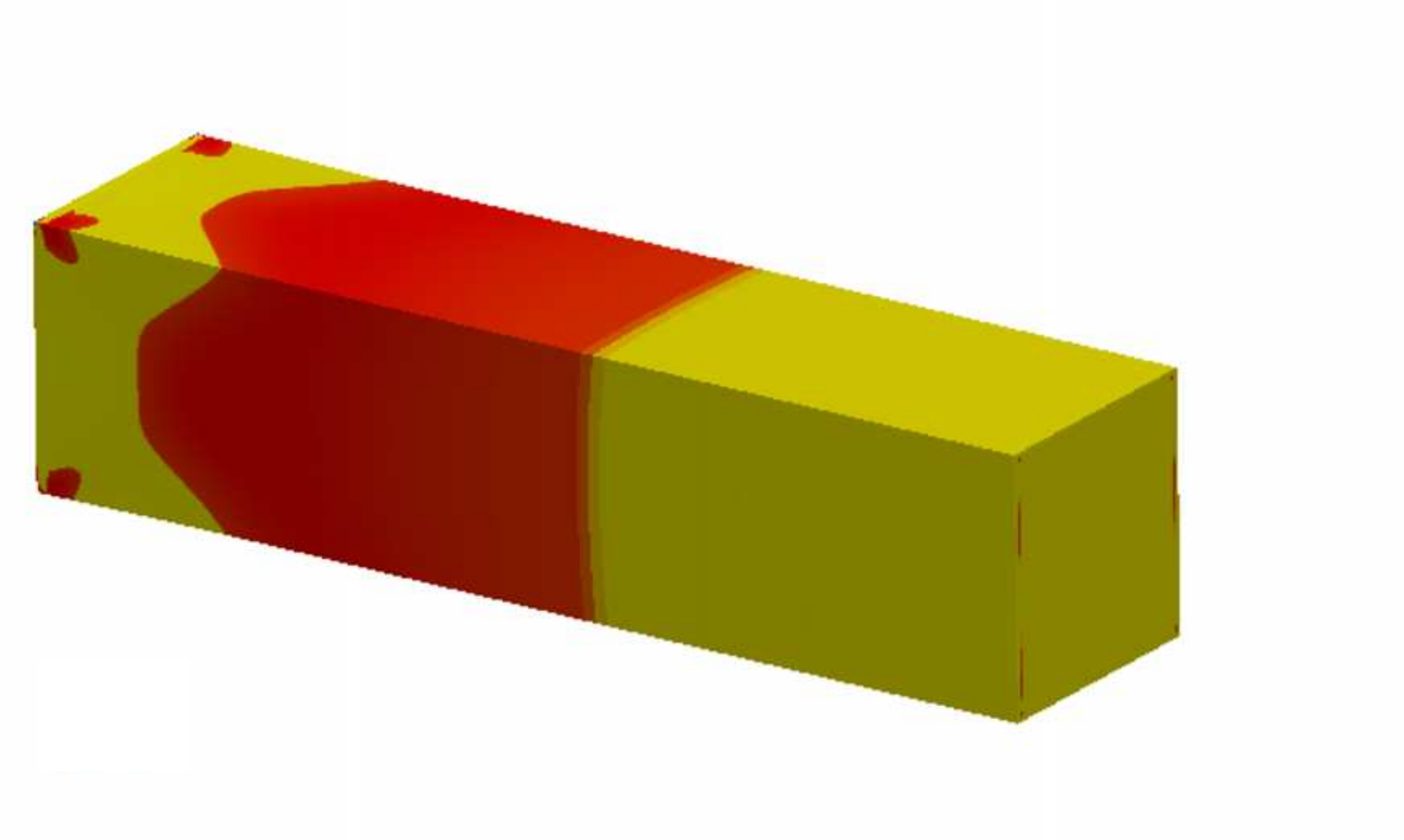}
\put(0,65){(c)}
\end{overpic}
}\\
\subfigure 
{
\begin{overpic}[trim=0mm 10mm 0mm 10mm,clip, width=0.3\linewidth] {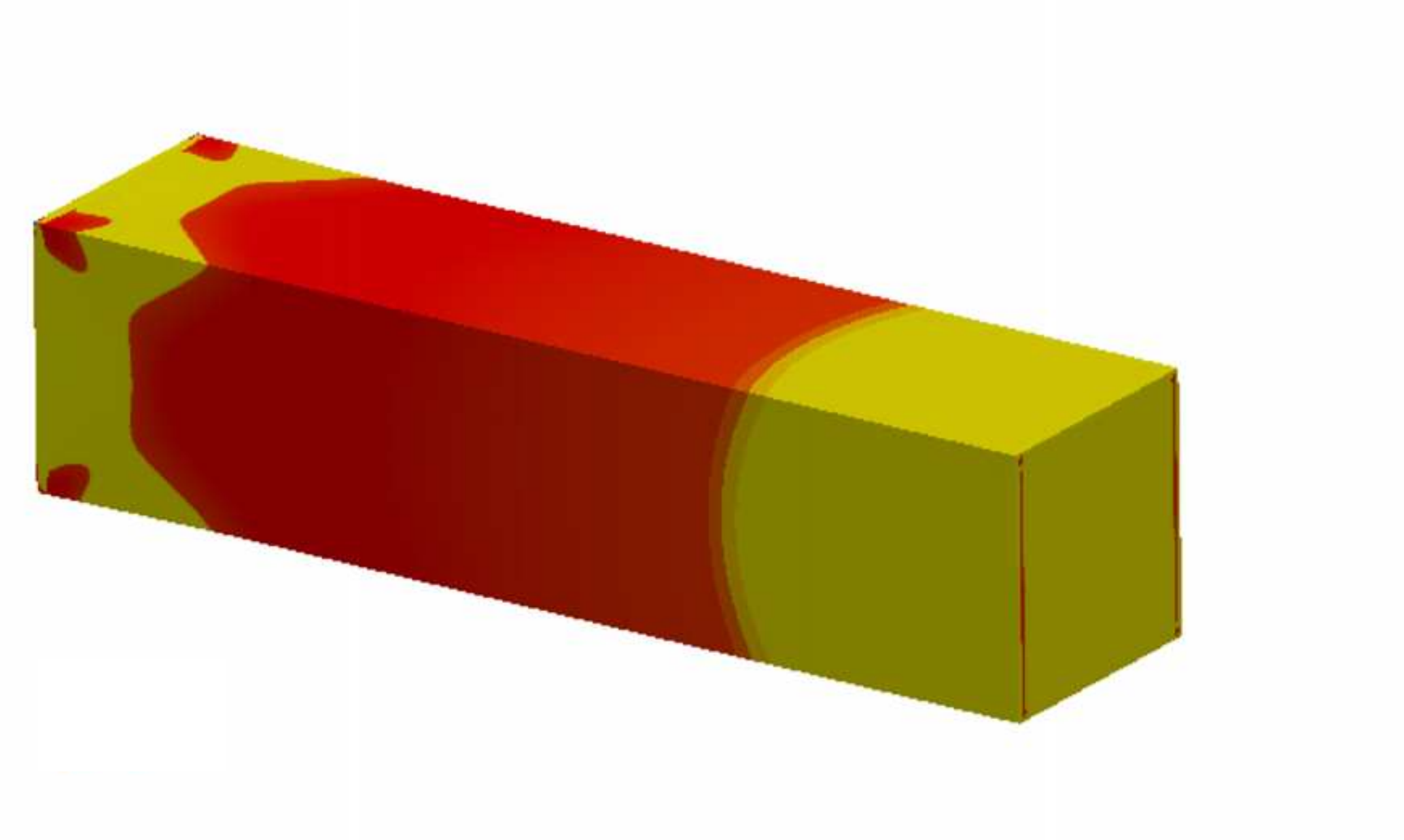}
\put(0,65){(d)}
\end{overpic}
}
\subfigure 
{
\begin{overpic}[trim=0mm 10mm 0mm 10mm,clip, width=0.3\linewidth] {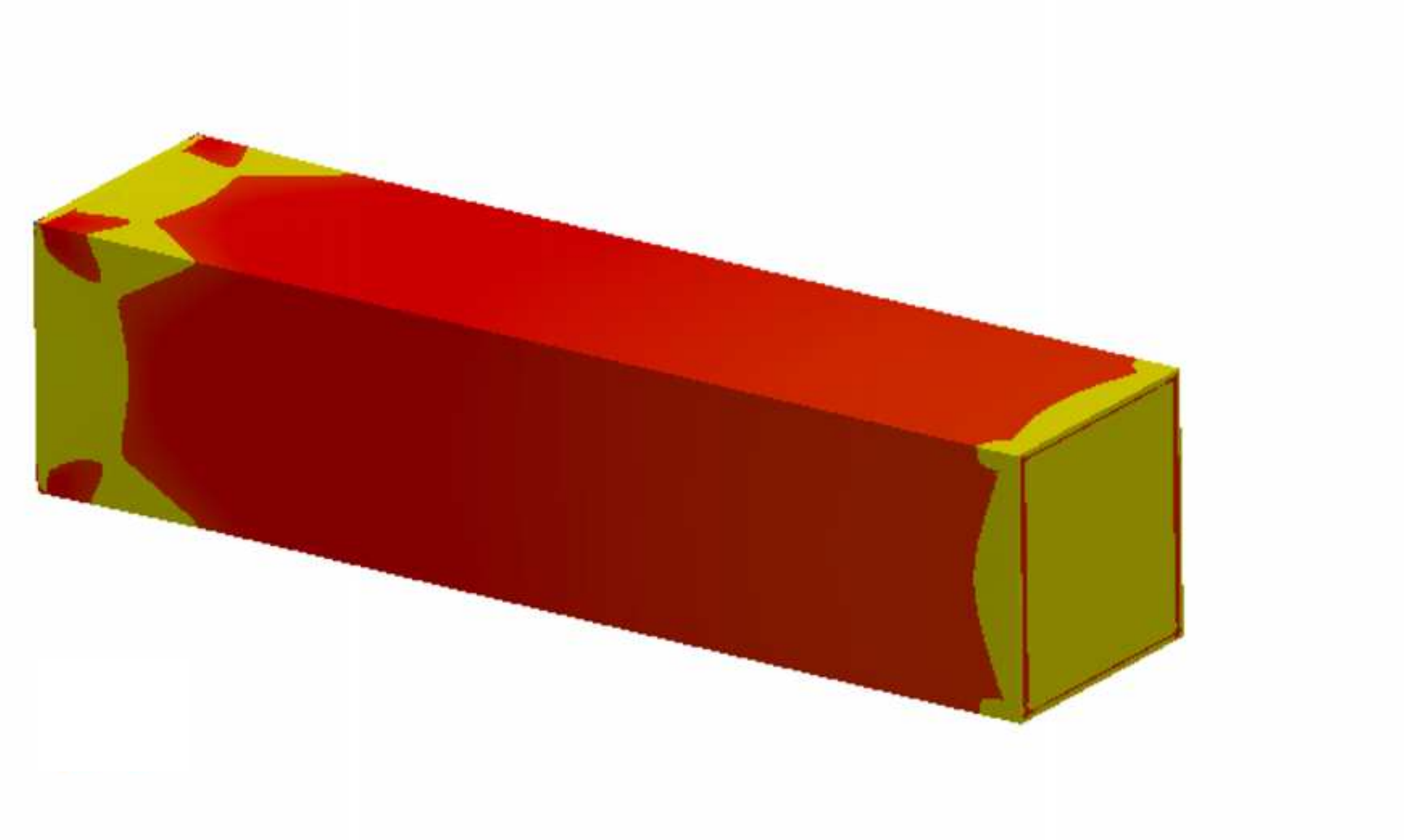}
\put(0,65){(e)}
\end{overpic}
}\\
\subfigure 
{
\begin{overpic}[trim=0mm 10mm 0mm 10mm,clip, width=0.3\linewidth] {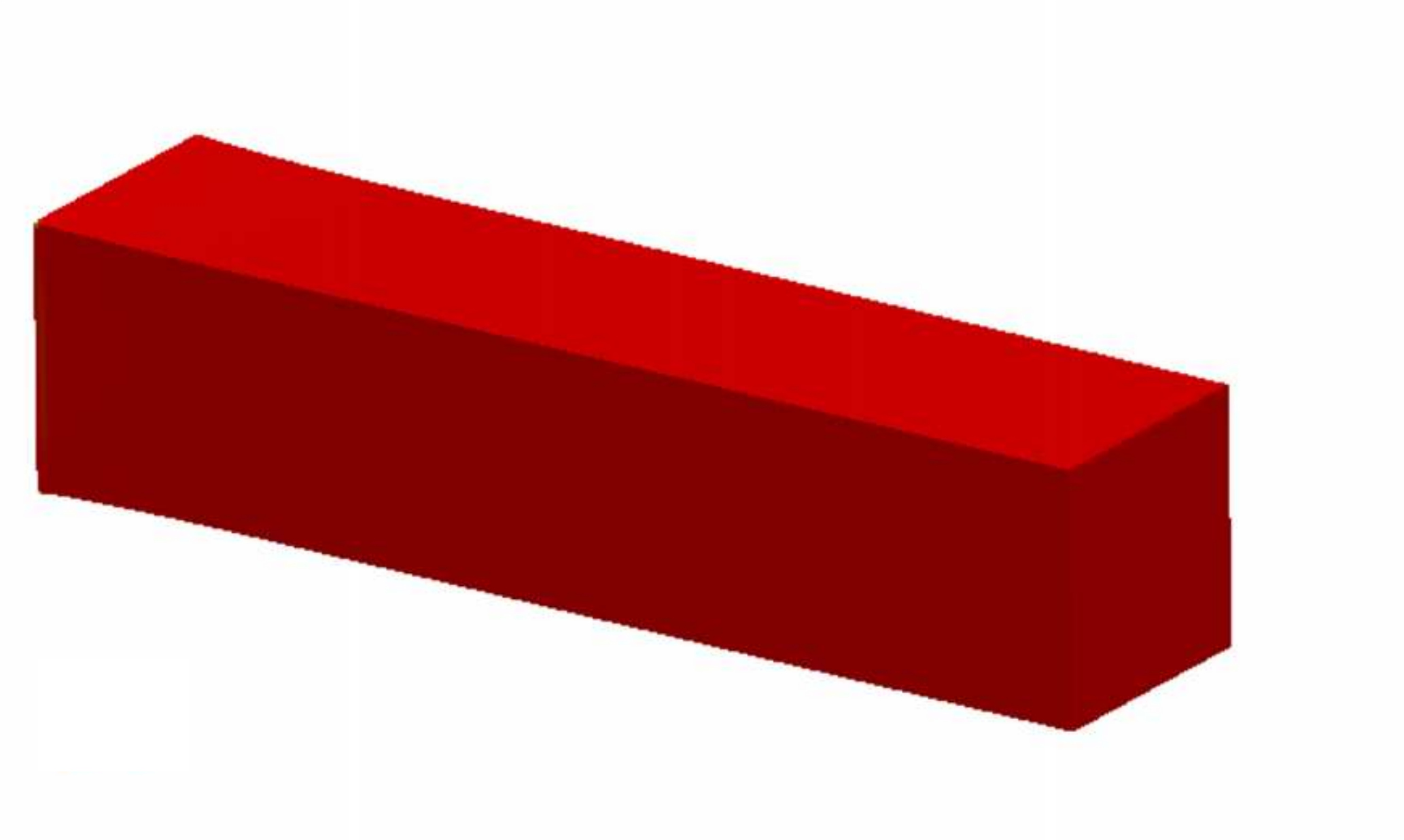}
\put(0,65){(f)}
\end{overpic}
}
\subfigure 
{
\begin{overpic}[trim=0mm 0mm 0mm 0mm,clip, width=0.3\linewidth] {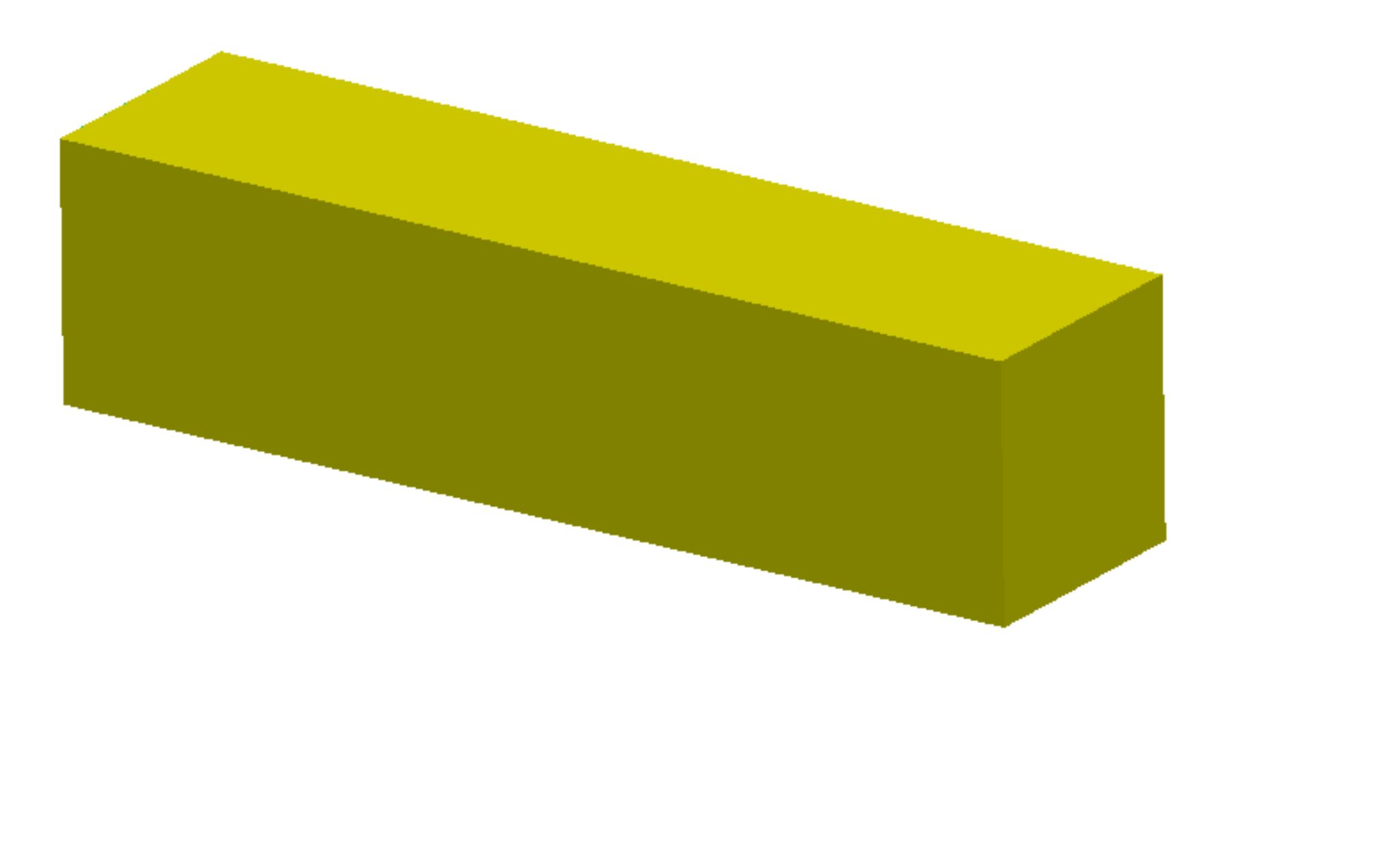}
\put(0,65){(g)}
\end{overpic}
}\\
\subfigure 
{
\begin{overpic}[trim=0mm 0mm 0mm 0mm,clip, width=0.3\linewidth]{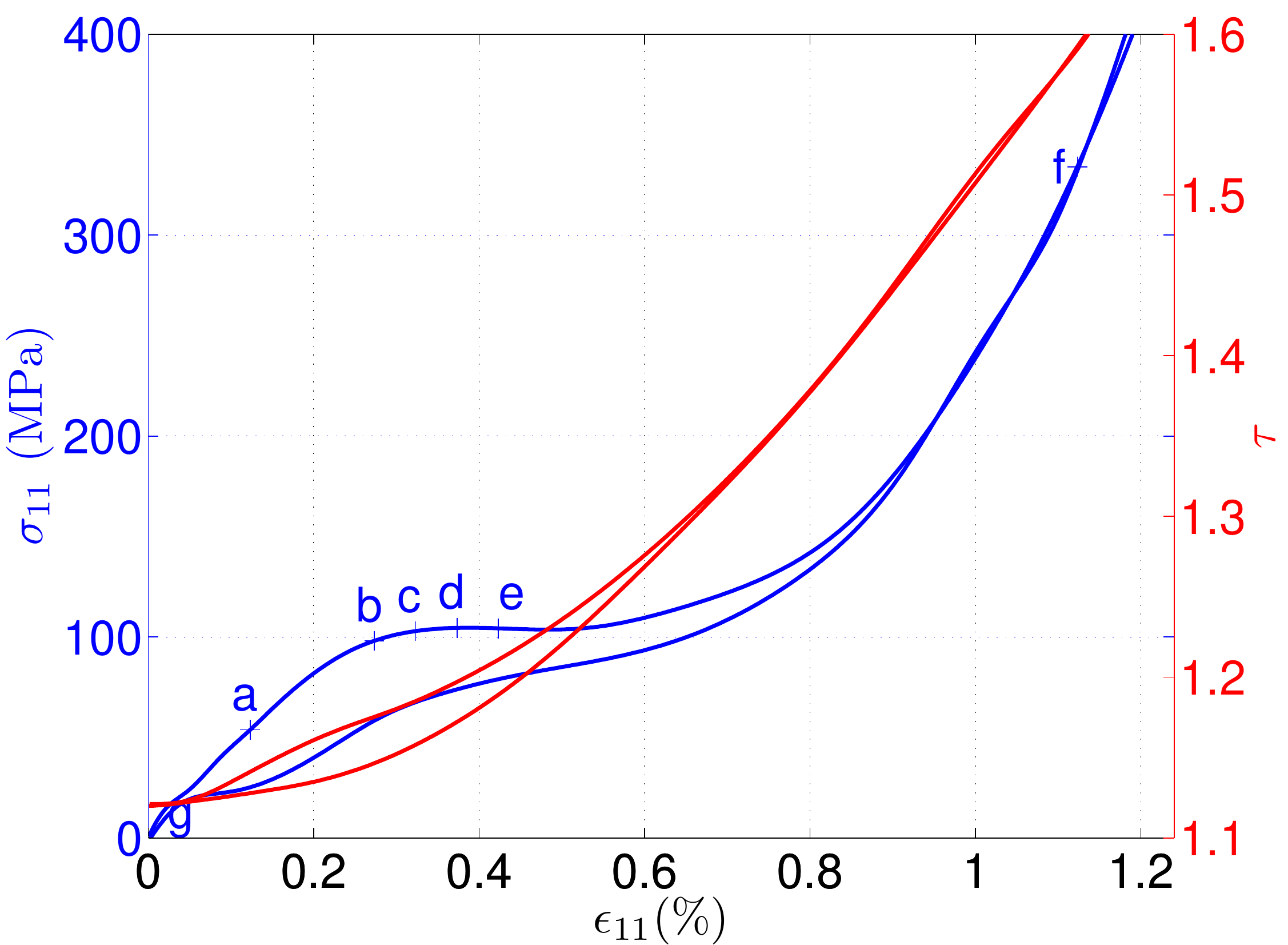}
\label{fig:PESxxVsExxVsTau}
\put(12,90){(h)}
\end{overpic}
}
\subfigure
{
\begin{overpic}[trim=0mm 0mm 0mm 0mm,clip, width=0.3\linewidth] {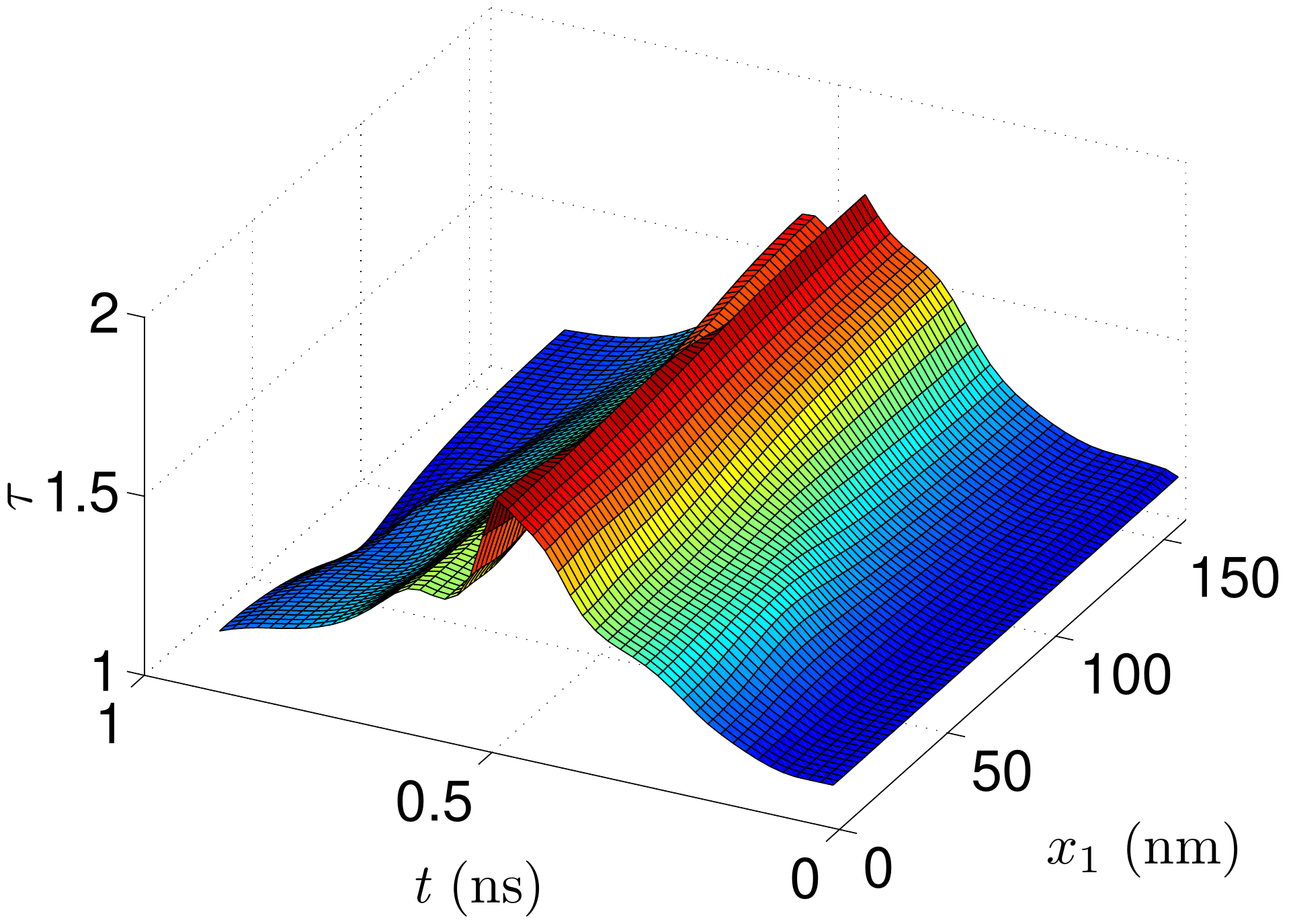}
\label{fig:PEExtrudeTau}
\put(12,90){(i)}
\put(100,85){\includegraphics[scale=.06]
{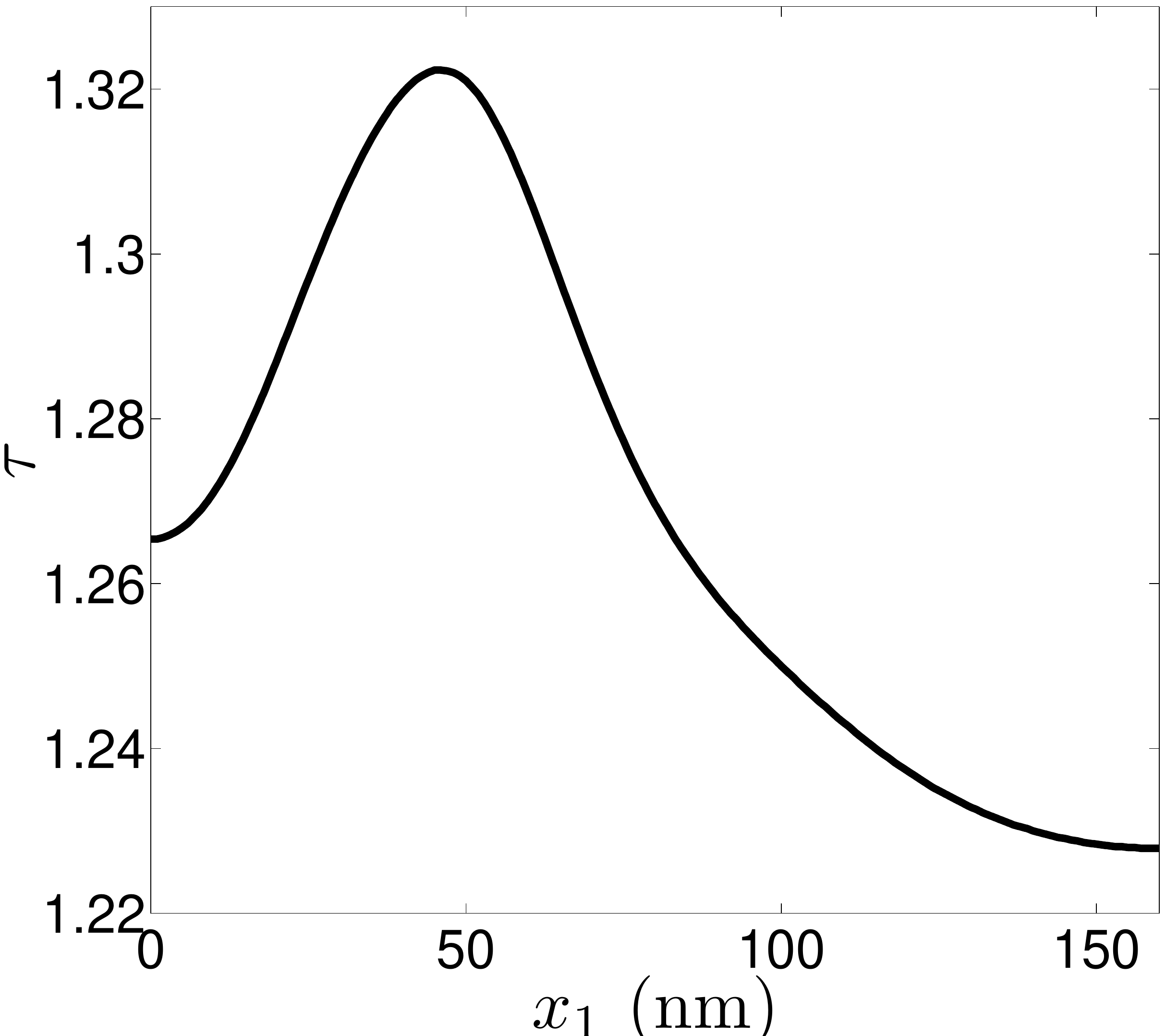}}
\put(105,90){{\tiny \textit{t}=0.225 ns}}
\end{overpic}
}
\caption{(Color online) Pseudoelasticity: Subplots (a-g) show the microstructure morphology evolution during loading and unloading (red and yellow colors represent M$_1$ variant  and austenite phase). Hysteresis plots of average $ \sigma_{11}-\epsilon_{11} $ (blue) and $\tau-\epsilon_{11} $ (red) are shown in subplot (h). Time extrusion plot of average $ \tau $ over the arc-length along the center line of the rectangular prism (between the opposite ends (0,20,20)--(160,20,20) nm) is shown in subplot (i). The schematic of loading-unloading cycle is shown in subplot (o).}
\label{fig:PEEvolution}
\end{figure}

To study the PE behavior, the SMA nanowire is quenched to the dimensionless temperature corresponding to $ \tau $ = 1.12 and allowed to evolve starting from random initial displacements and periodic boundary conditions. The nanowire has been evolved till the microstructure and free energy are stabilized. The nanowire remains in the austenite phase. The evolved microstructures are then axially loaded by mechanically constraining one end of the specimen  $\vect u = \vect 0$, and loading the opposite end using a ramp based displacement equivalent to the strain rate \mbox{3$\times$10$^7/$s} as shown in Fig. \ref{fig:PEEvolution}(o). Figure \ref{fig:PEEvolution}(a-g) shows  different time snapshots of microstructure evolution during loading (a-f) and the end of unloading cycle (g). The austenite (yellow) is converted into the favorable \mOne martensite (red) with the formation and migration of habit plane in the nanowire. At the end of unloading, the nanowire returns to the austenite phase (refer to Fig. \ref{fig:PEEvolution}(g)). The mechanical hysteresis, the average $\sigma_{11}$--$\epsilon_{11} $ (blue color), forms a full loop with zero remnant strain at the end of unloading. The temperature hysteresis, the average $\tau$--$\epsilon_{11} $ (red color), indicates a global increase in the temperature during loading and a decrease during unloading as a result of exothermic and endothermic processes. 
 
The phase transformation is a local phenomenon that leads to non-uniform deformation and temperature fields in the domain. The local variation of the non-dimensional temperature $ \tau $ is presented as an arc-length extrusion plot along the central axis of the nanowire during loading and unloading in Fig. \ref{fig:PEEvolution}(i) . The  non-uniform strain and deformation are apparent during phase transformations. The local increase in temperature, as observed in Figs. \ref{fig:PEEvolution}(i) serves as a signature of formation or movement of habit planes as indicated in the inset. As the loading progresses, the heat produced is conducted in the domain causing self-heating thus increasing the global temperature. As the domain is small, the heat is conducted quickly causing small local peaks as compared to the experiments where large temperature peaks were observed in big macro specimens \cite{shaw1995thermomechanical}.

\begin{figure}[h!]
\centering
\subfigure 
{
\begin{overpic}[trim=0mm 10mm 0mm 10mm,clip, width=0.3\linewidth] {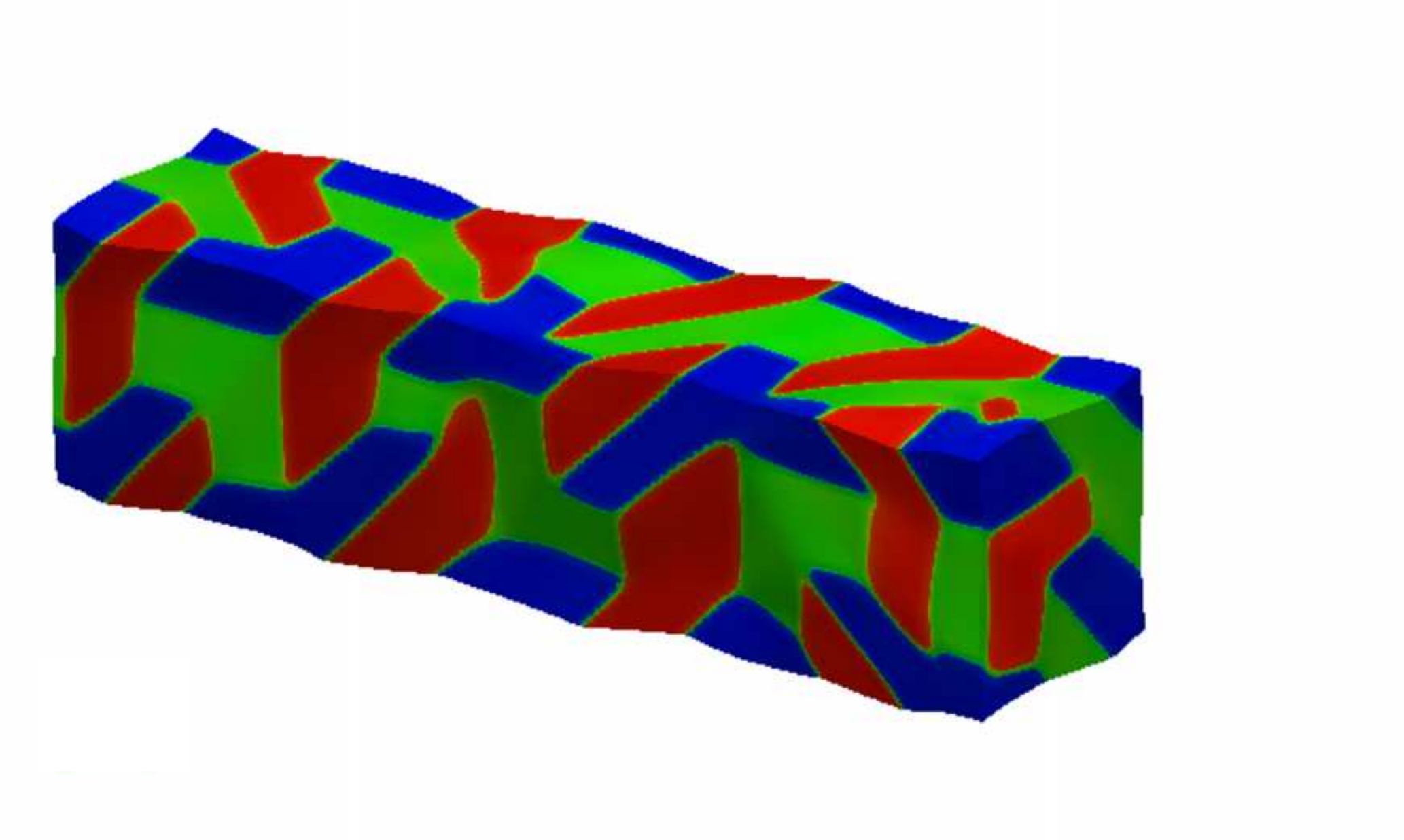}
\put(0,65){(a)}
\end{overpic}
}
\subfigure 
{
\begin{overpic}[trim=0mm 10mm 0mm 10mm,clip, width=0.3\linewidth] {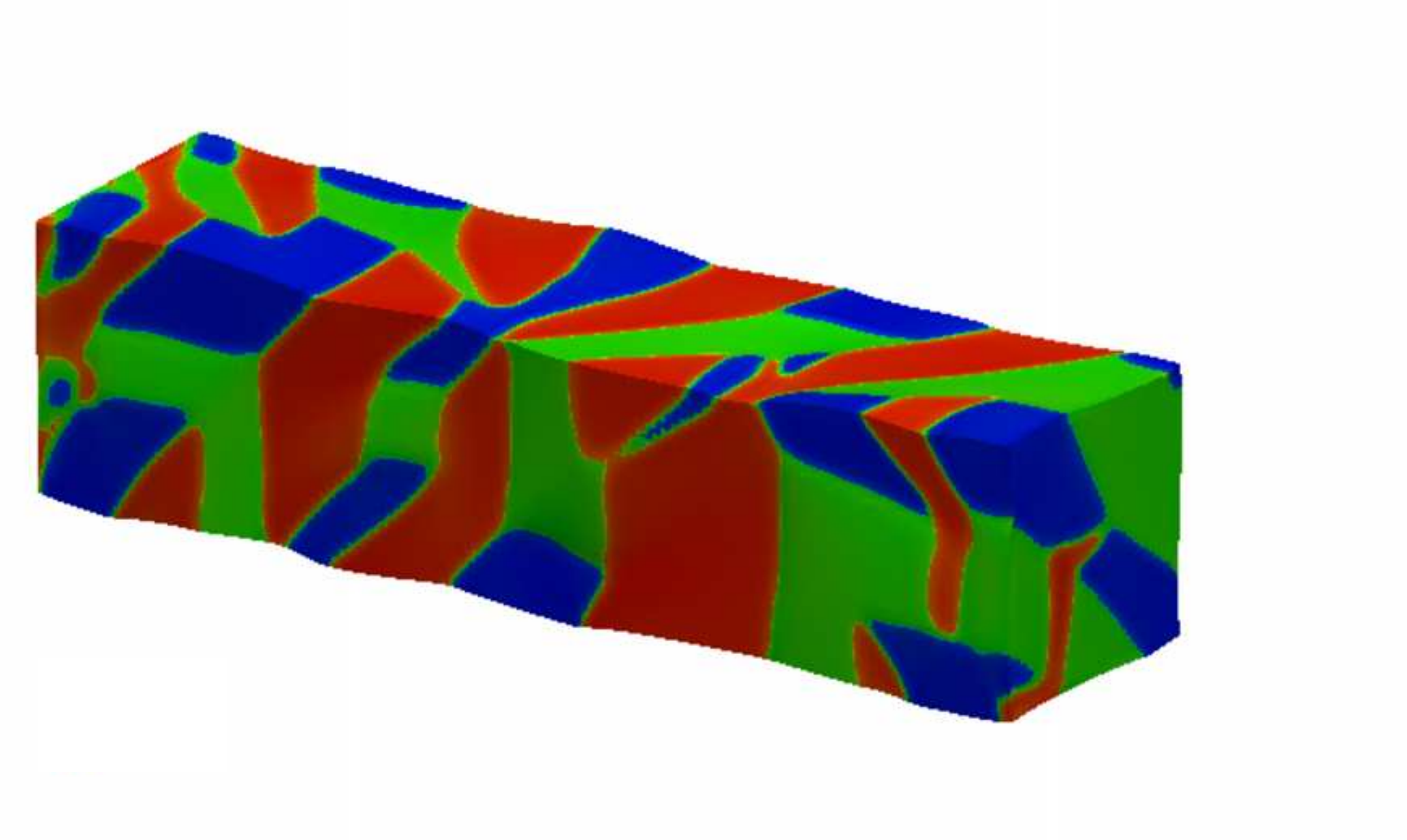}
\put(0,65){(b)}
\end{overpic}
}\\
\subfigure 
{
\begin{overpic}[trim=0mm 10mm 0mm 10mm,clip, width=0.3\linewidth] {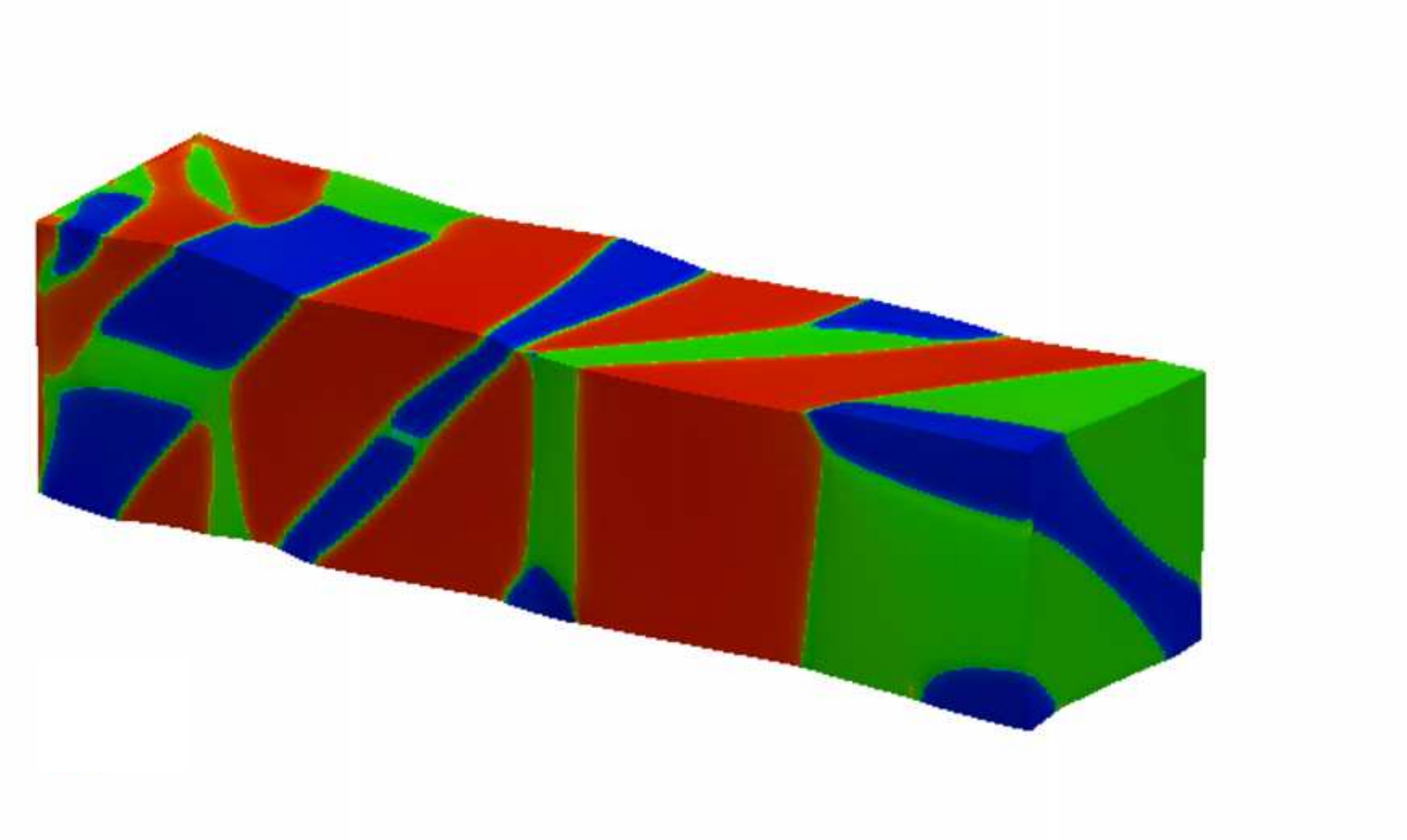}
\put(0,65){(c)}
\end{overpic}
}
\subfigure 
{
\begin{overpic}[trim=0mm 10mm 0mm 10mm,clip, width=0.3\linewidth] {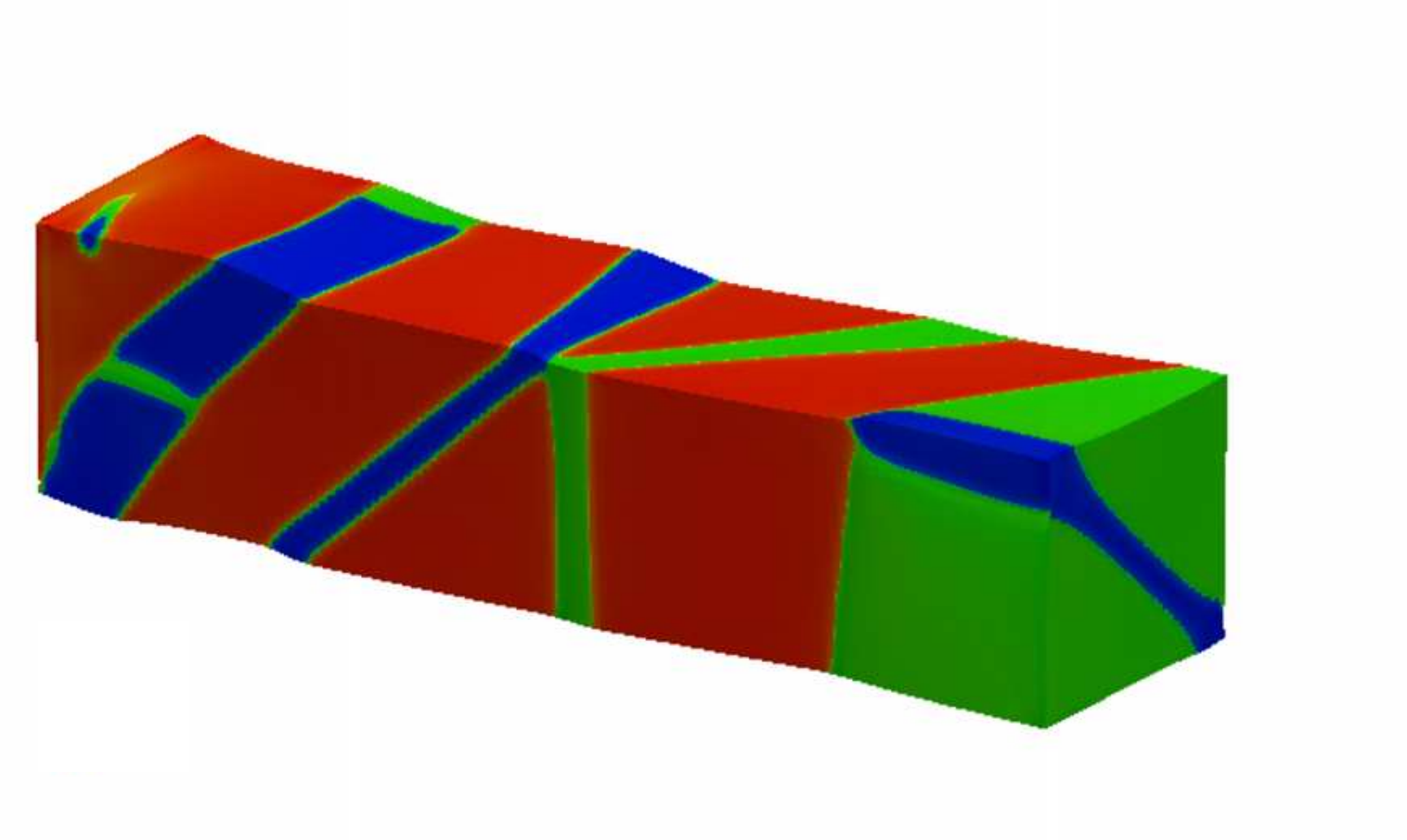}
\put(0,65){(d)}
\end{overpic}
}\\
\subfigure 
{
\begin{overpic}[trim=0mm 10mm 0mm 10mm,clip, width=0.3\linewidth] {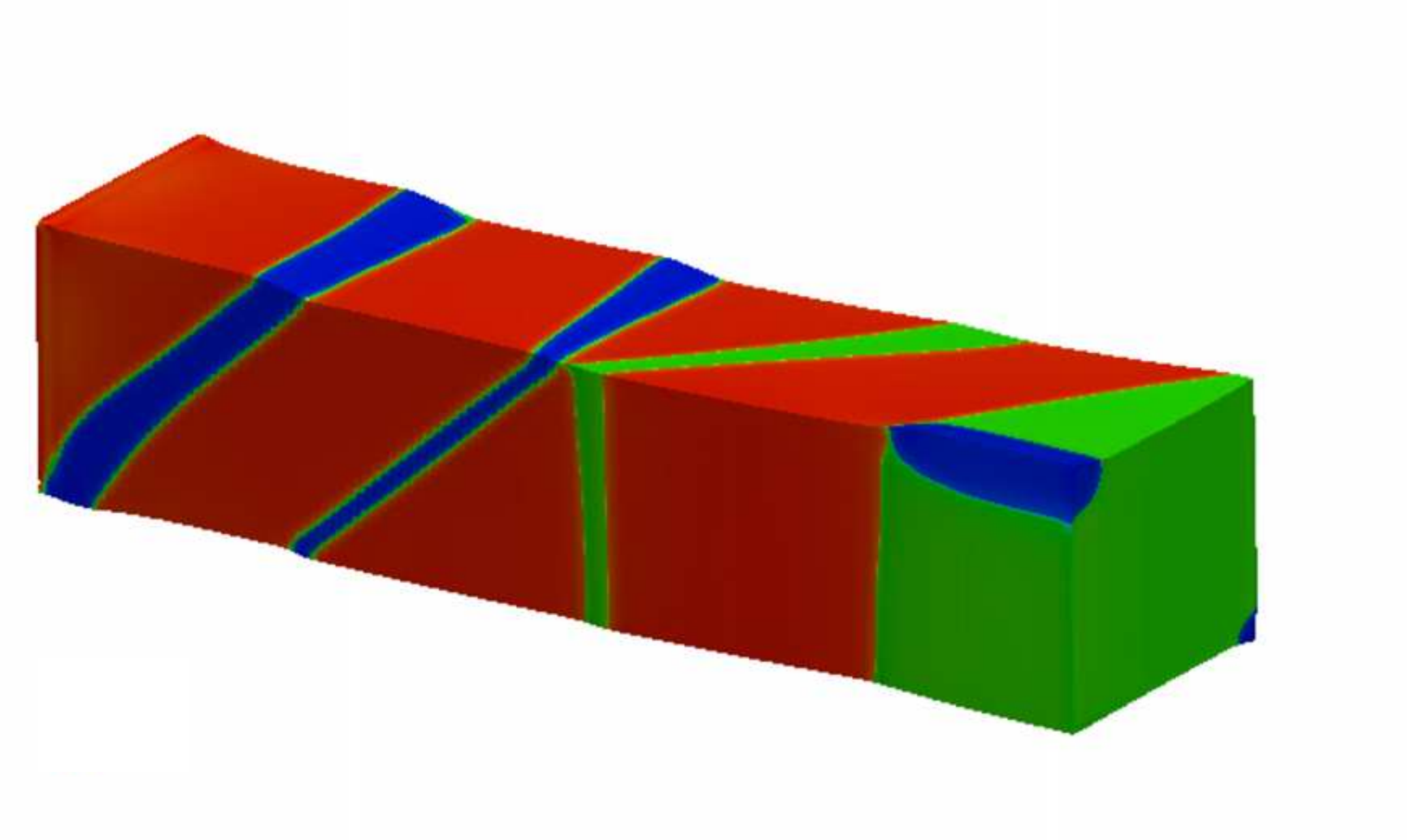}
\put(0,65){(e)}
\end{overpic}
}
\subfigure 
{
\begin{overpic}[trim=0mm 10mm 0mm 10mm,clip, width=0.3\linewidth] {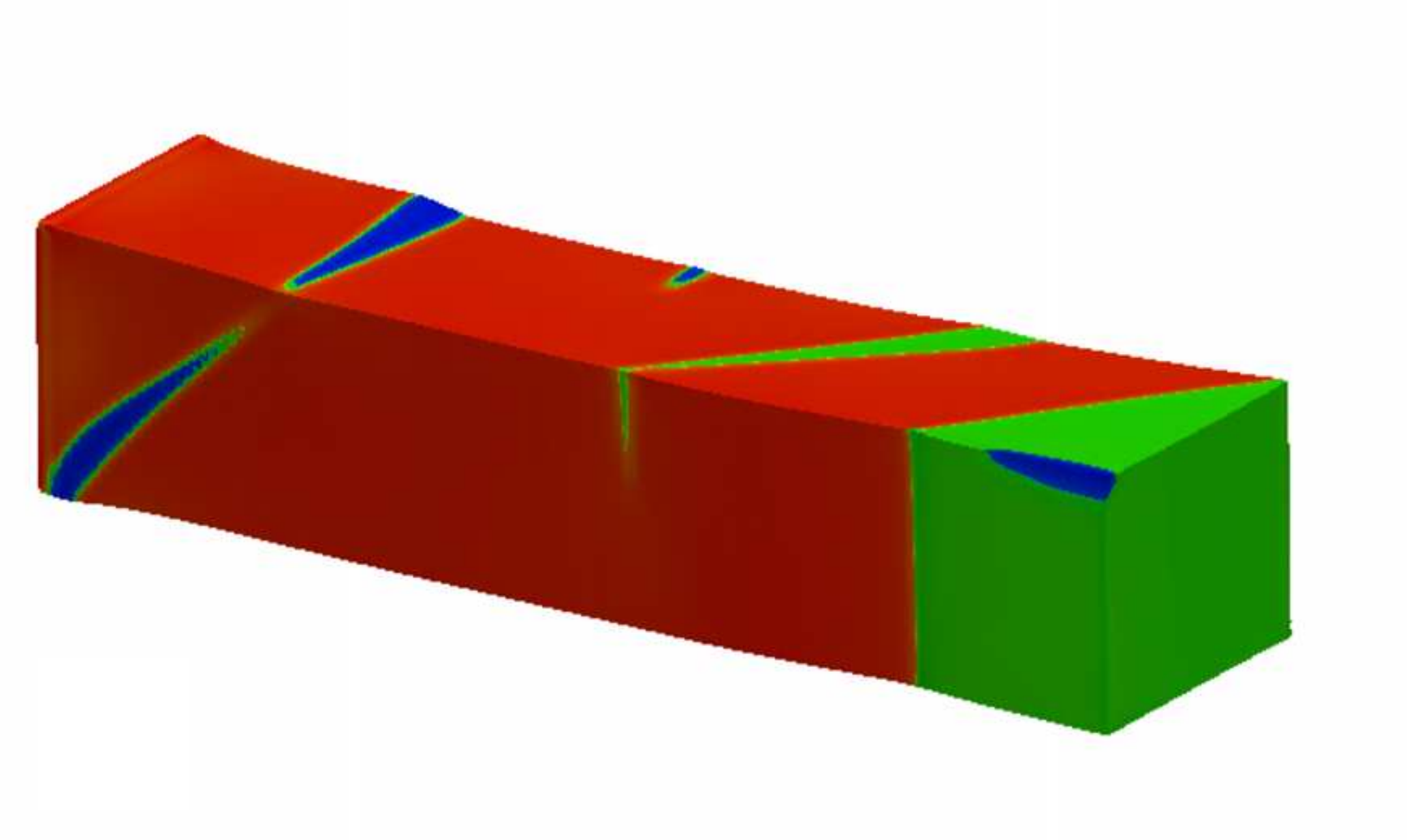}
\put(0,65){(f)}
\end{overpic}
}\\
\subfigure 
{
\begin{overpic}[trim=0mm 10mm 0mm 10mm,clip, width=0.3\linewidth] {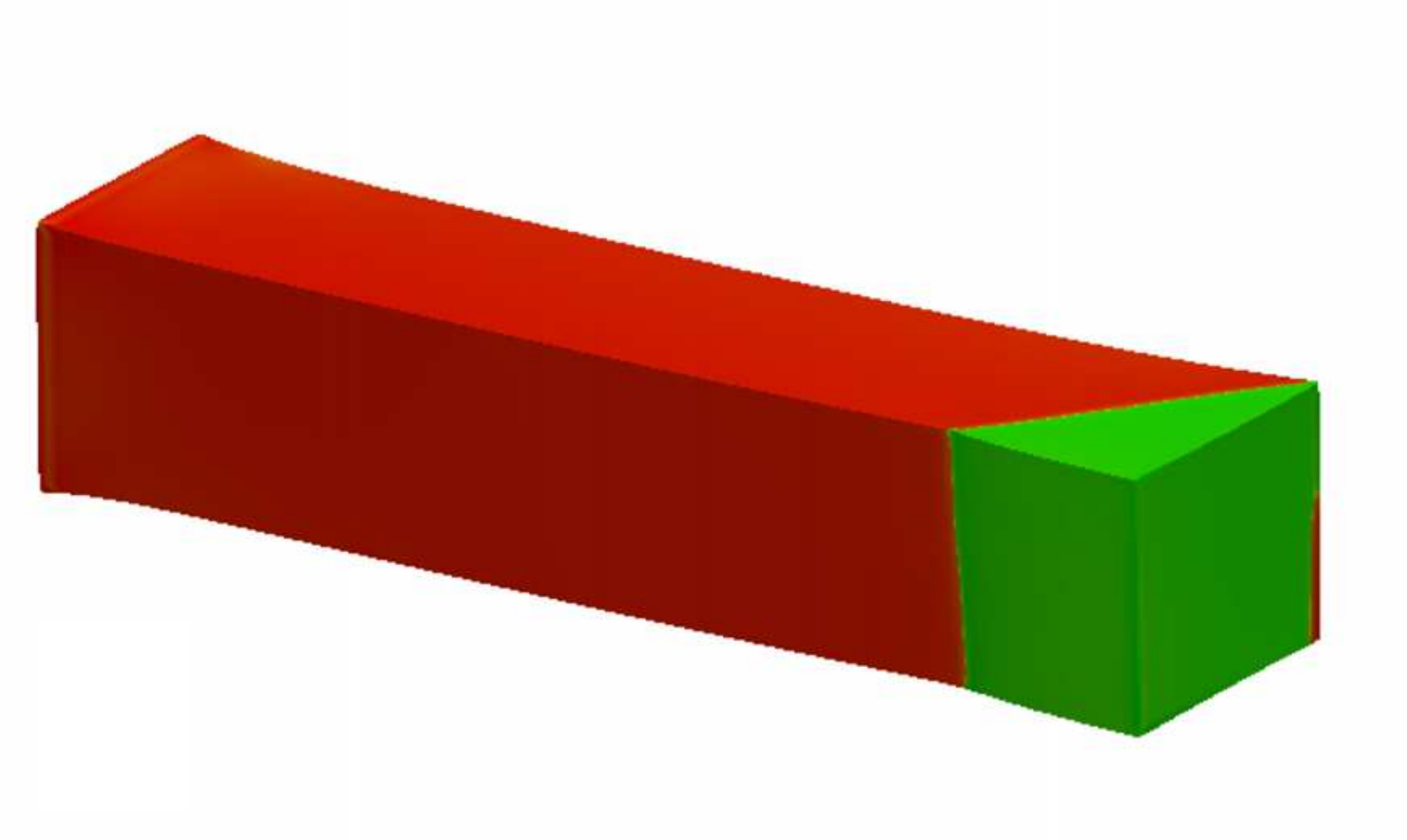}
\put(0,65){(g)}
\end{overpic}
}
\subfigure 
{
\begin{overpic}[trim=0mm 10mm 0mm 10mm,clip, width=0.3\linewidth] {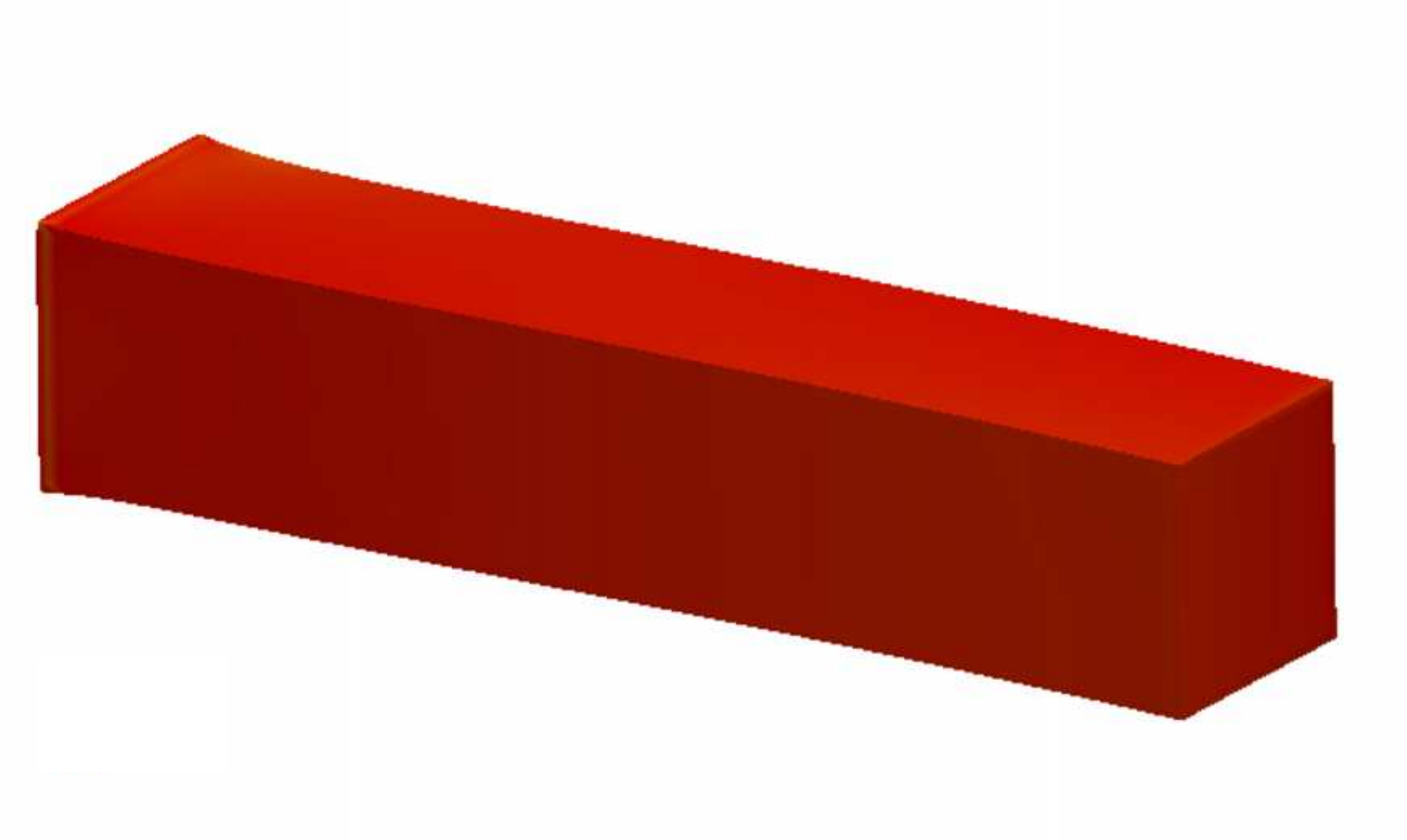}
\put(0,65){(h)}
\end{overpic}
}\\
\subfigure 
{
\begin{overpic}[trim=0mm 0mm 0mm 0mm,clip, width=0.3\linewidth] {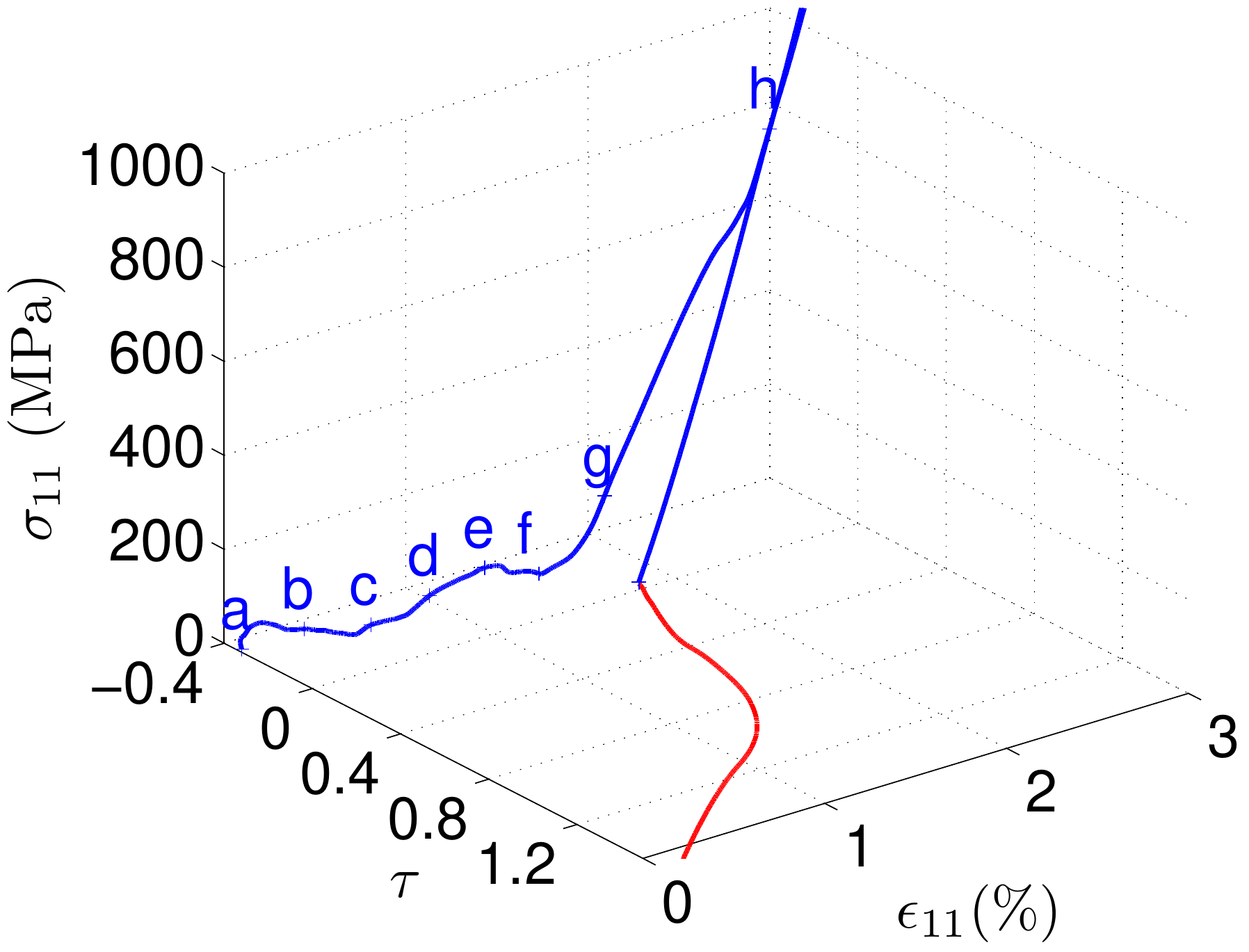}
\label{fig:SMESxxVsExxVsTau}
\put(25,90){(i)}
\end{overpic}
}
\subfigure
{
\begin{overpic}[trim=0mm 0mm 0mm 0mm,clip, width=0.3\linewidth] {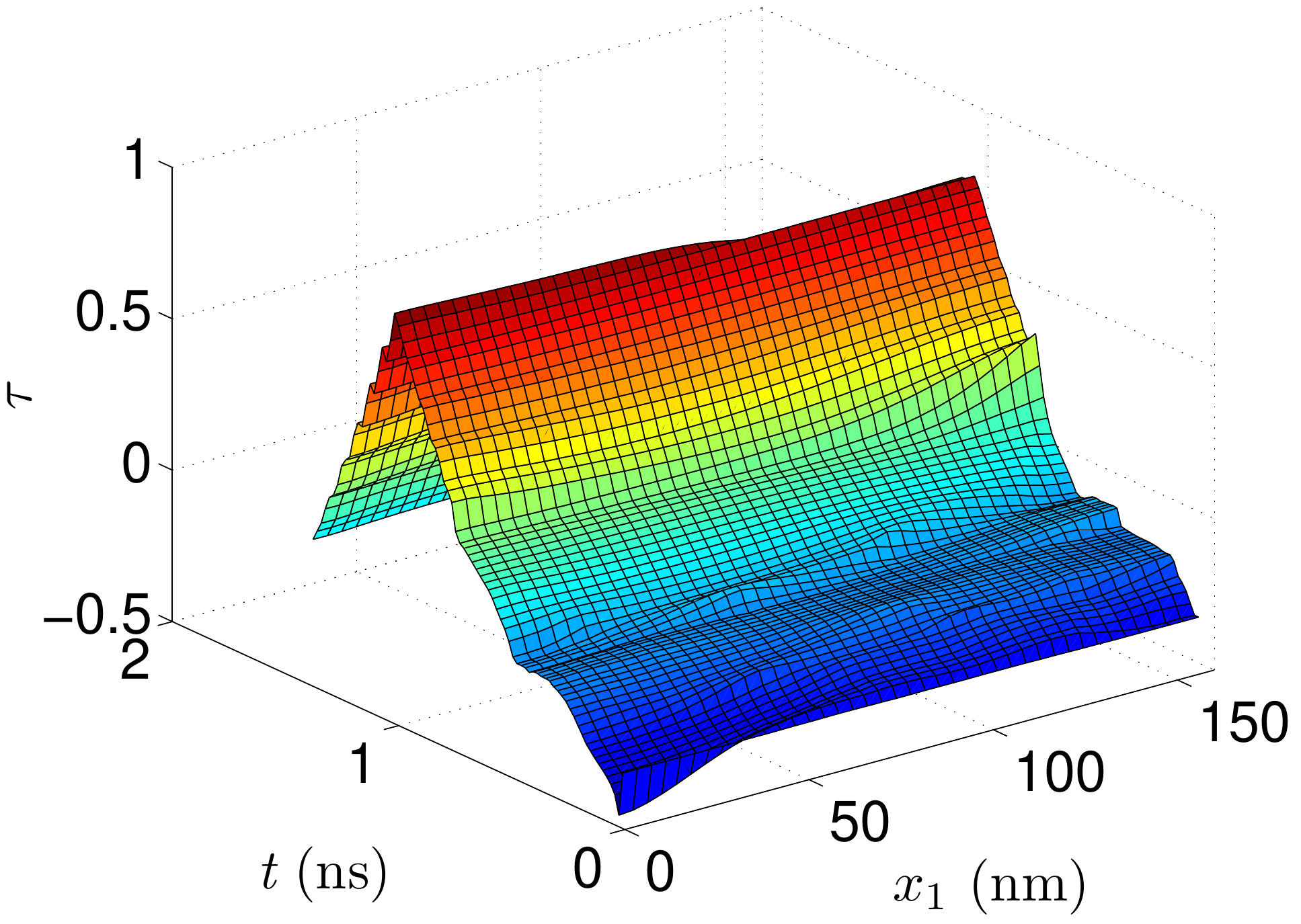}
\label{fig:SMEExtrudeTau}
\put(25,90){(j)}
\put(120,65){\includegraphics[scale=.055]
{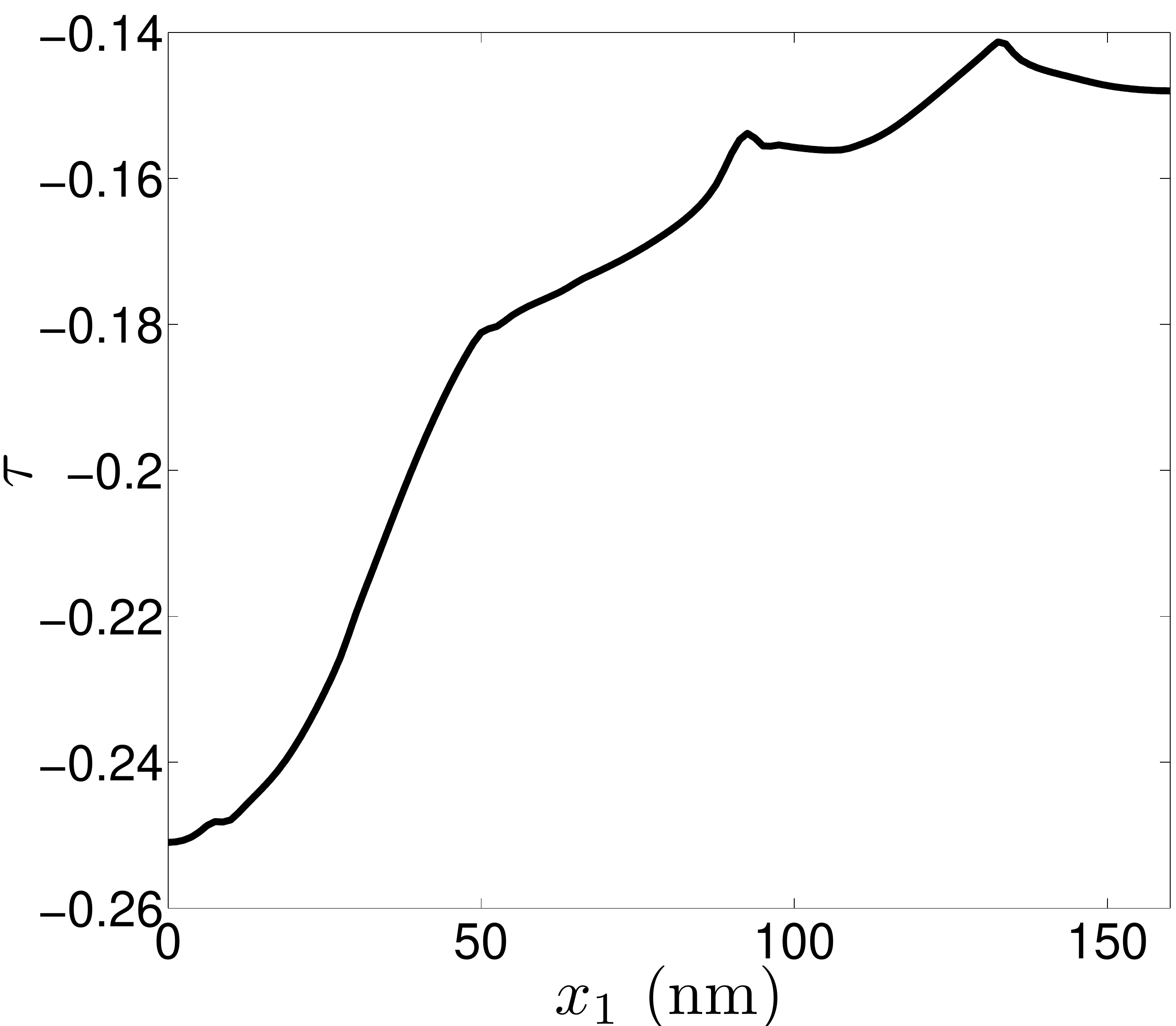}}
\put(130,70){{\tiny \textit{t}=0.189 ns}}
\end{overpic}
}
\caption{(Color online) Shape memory effect: Subplots (a-h) show the microstructure morphology evolution during loading and unloading (red, blue, and green colors represent M$_1$, M$_2$, and M$_3$ variants, respectively). The full SME loop behavior is shown in subplot (h) (red color indicates heating part of the cycle) . Time extrusion plot of average $ \tau $ over the arc-length along the center line of the rectangular prism (between the opposite ends (0,20,20)--(160,20,20) nm) is shown in subplot (i).}
\label{fig:SMEEvolution}
\end{figure}

To study the SME behavior, the SMA nanowire is quenched to the temperature corresponding to $ \tau = -1.2$ and allowing the microstructures and energy to stabilize. The nanowire domain is evolved into the accommodated twinned martensites with all three variants present in approximately equal proportions as shown in Fig. \ref{fig:SMEEvolution}(a). Next, the evolved microstructures are axially loaded by mechanically constraining one end of the specimen, and loading the opposite end using a ramp based displacement equivalent to the strain rate \mbox{3 $\times$ 10$^7/$s}.  The movement of domain walls, merging and elimination of unfavorable domains and growth of favorable ones are apparent.  During loading, the accommodated twinned microstructures are transformed into a detwinned microstructure via a phase transformation of unfavorable \mTwo (blue) and \mThree (green) variants to the favorable \mOne martensite (red). The domain walls migrate along \pooz planes. On unloading, the nanowire exists in \mOne phase at axial stress in the domain being zero (refer to Fig. \ref{fig:SMEEvolution}(h)). The nanowire is then unloaded to zero stress-strain configuration with the same strain rate but now by applying an external thermal load $g$ = 0.1 (in the dimensionless units). The nanowire is converted to the austenite phase forming a full-SME loop. 

The local variation of non-dimensional temperature  $ \tau $ is presented as an arc-length extrusion plot in Fig. \ref{fig:SMEEvolution}(j). Non-uniform strain and deformation are apparent during phase transformations. The local deformation is non-uniform during loading and it becomes nearly uniform after all the unfavorable variants are converted to the favorable ones. The local increase in temperature at one time instant is indicated in the inset. As the loading progresses, the heat produced is conducted in the domain causing self-heating thus increasing the global temperature. The model qualitatively captures the local phenomenon observed experimentally \cite{pieczyska2006phase,shaw1997nucleation}.

The developed model has enabled us to capture the inherent temperature hysteresis qualitatively, which is in agreement with the experiments \cite{pieczyska2006phase}. The model has also successfully captured  thermal hysteresis $\tau$--$\epsilon_{11}$ in the PE and the complete $\sigma_{11}$--$\epsilon_{11} $--$\tau$ loop in SME. Under the dynamic loading conditions of SMA nanowires, this has been done for the first time. 

\section{Nanoscale Dynamics of Springs}  
\begin{figure}[h!]
\centering
\subfigure
{
\begin{overpic}[trim=35mm 5mm 35mm 35mm,clip, width=0.12\linewidth]  {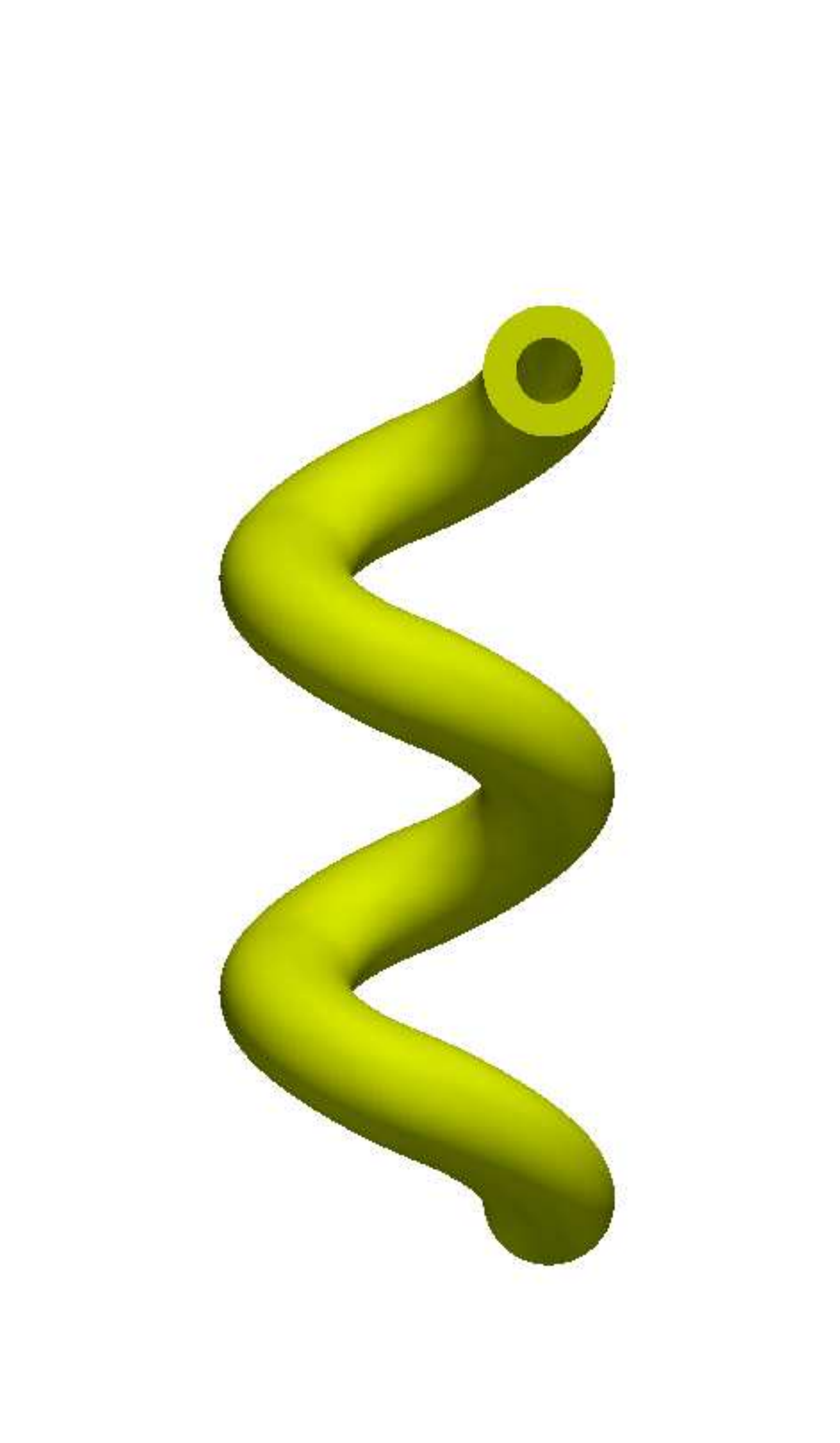}
\put(0,130){(a)}
\end{overpic}
}
\subfigure
{
\begin{overpic}[trim=35mm 5mm 35mm 35mm,clip, width=0.12\linewidth]  {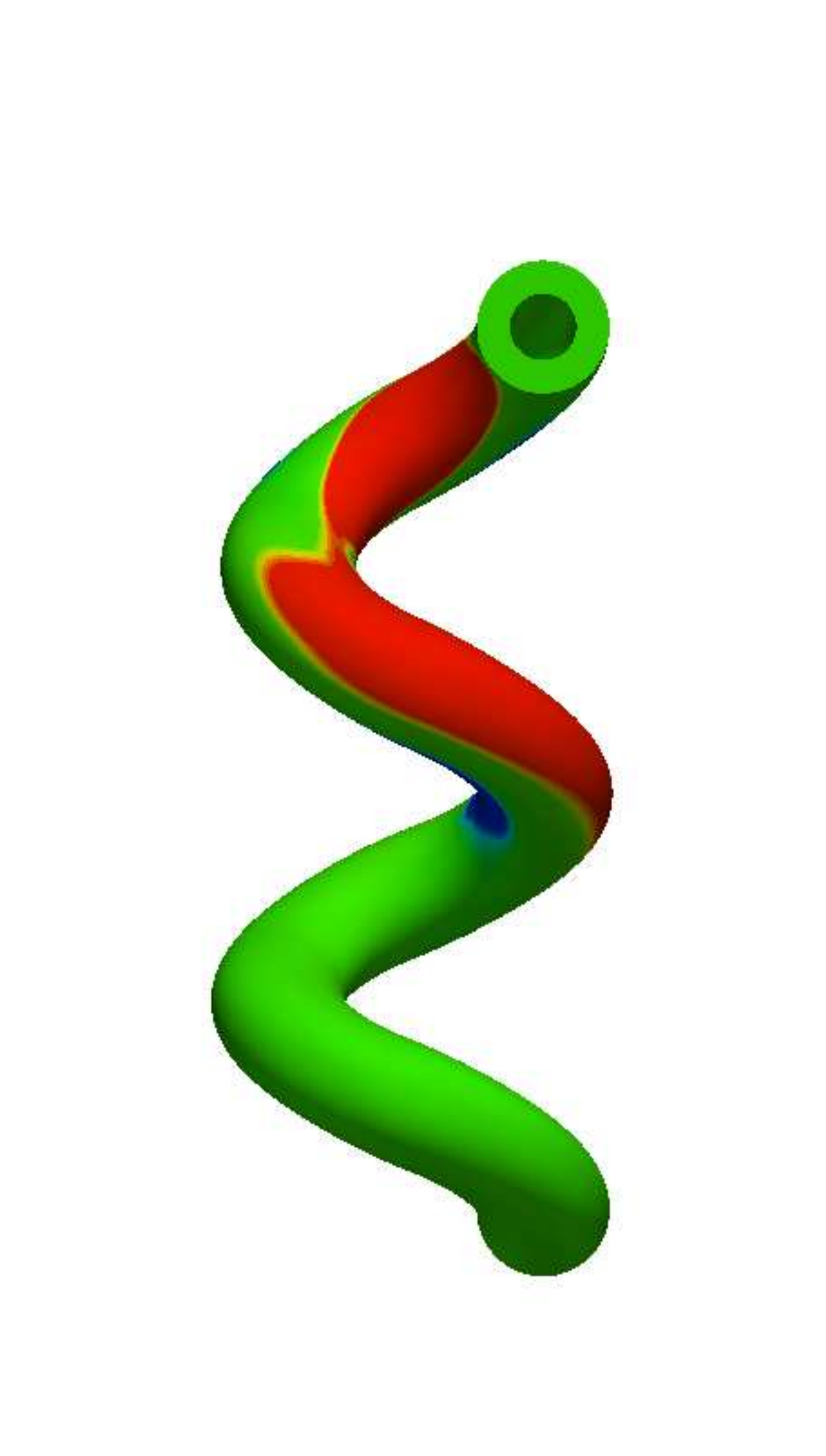}
\put(0,130){(b)}
\end{overpic}
}
\subfigure
{
\begin{overpic}[trim=35mm 5mm 35mm 35mm,clip, width=0.12\linewidth]  {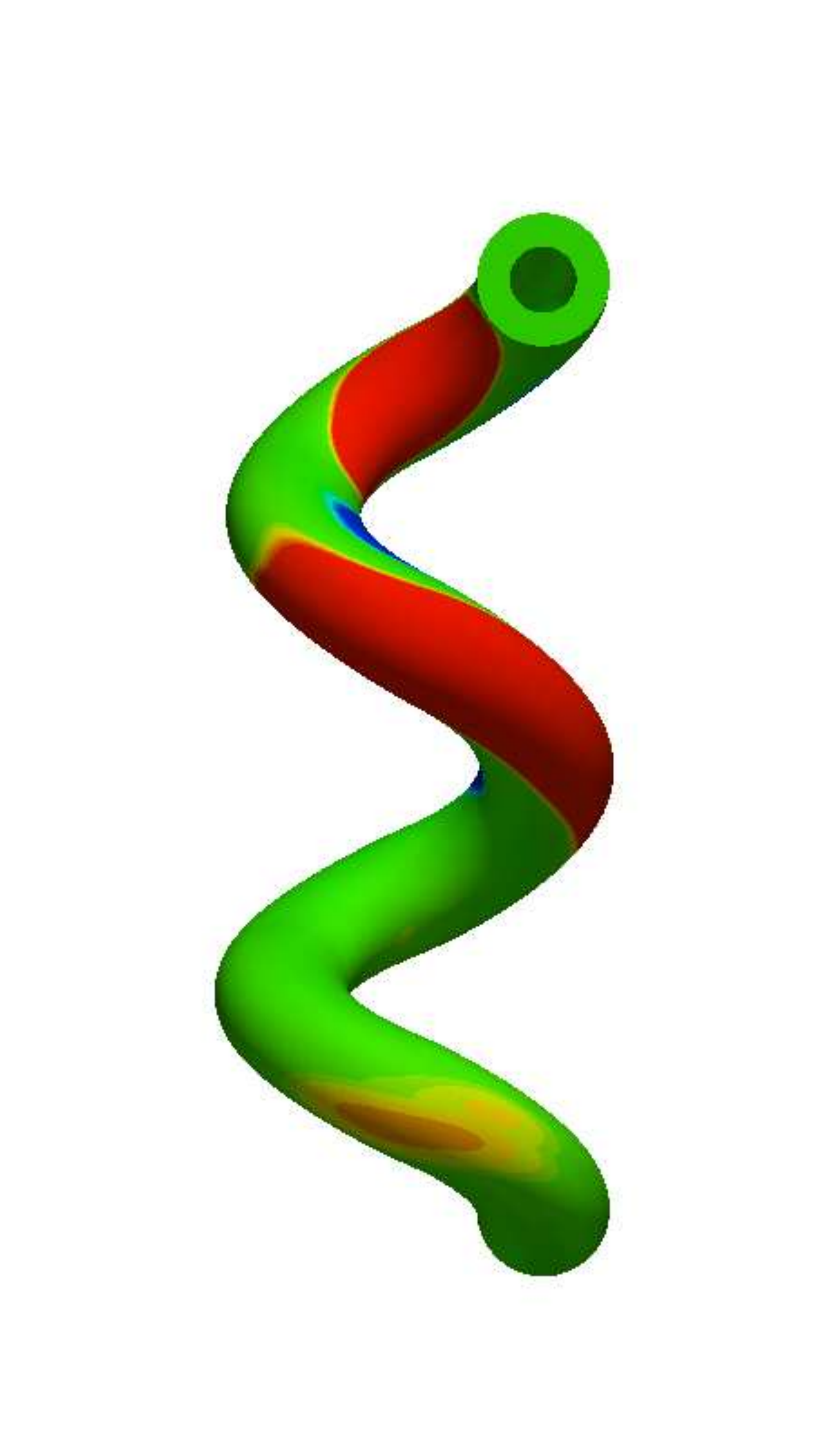}
\put(0,130){(c)}
\end{overpic}
}
\subfigure
{
\begin{overpic}[trim=35mm 5mm 35mm 35mm,clip, width=0.12\linewidth]  {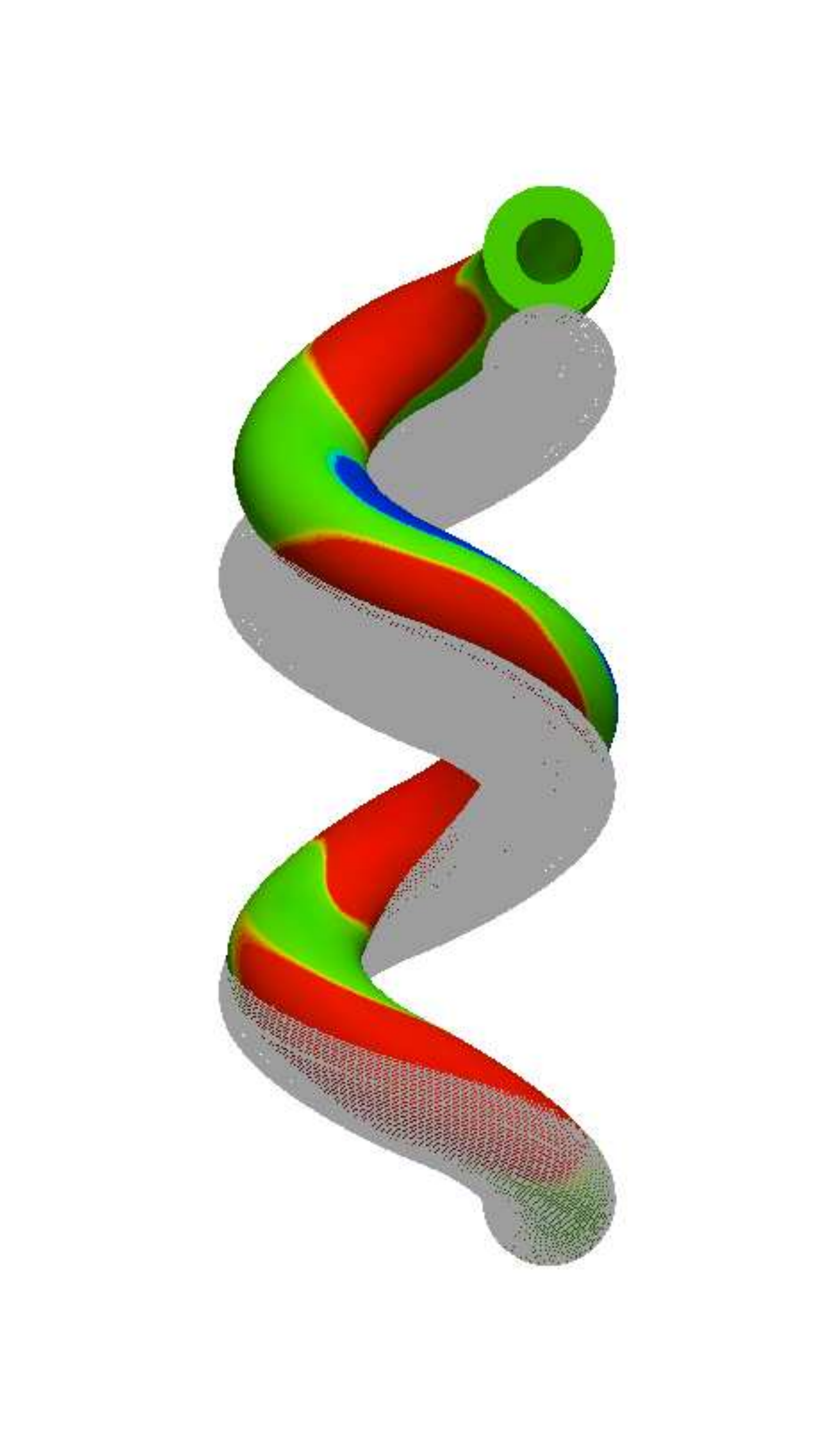}
\put(0,130){(d)}
\end{overpic}
}
\subfigure
{
\begin{overpic}[trim=35mm 5mm 35mm 35mm,clip, width=0.12\linewidth]  {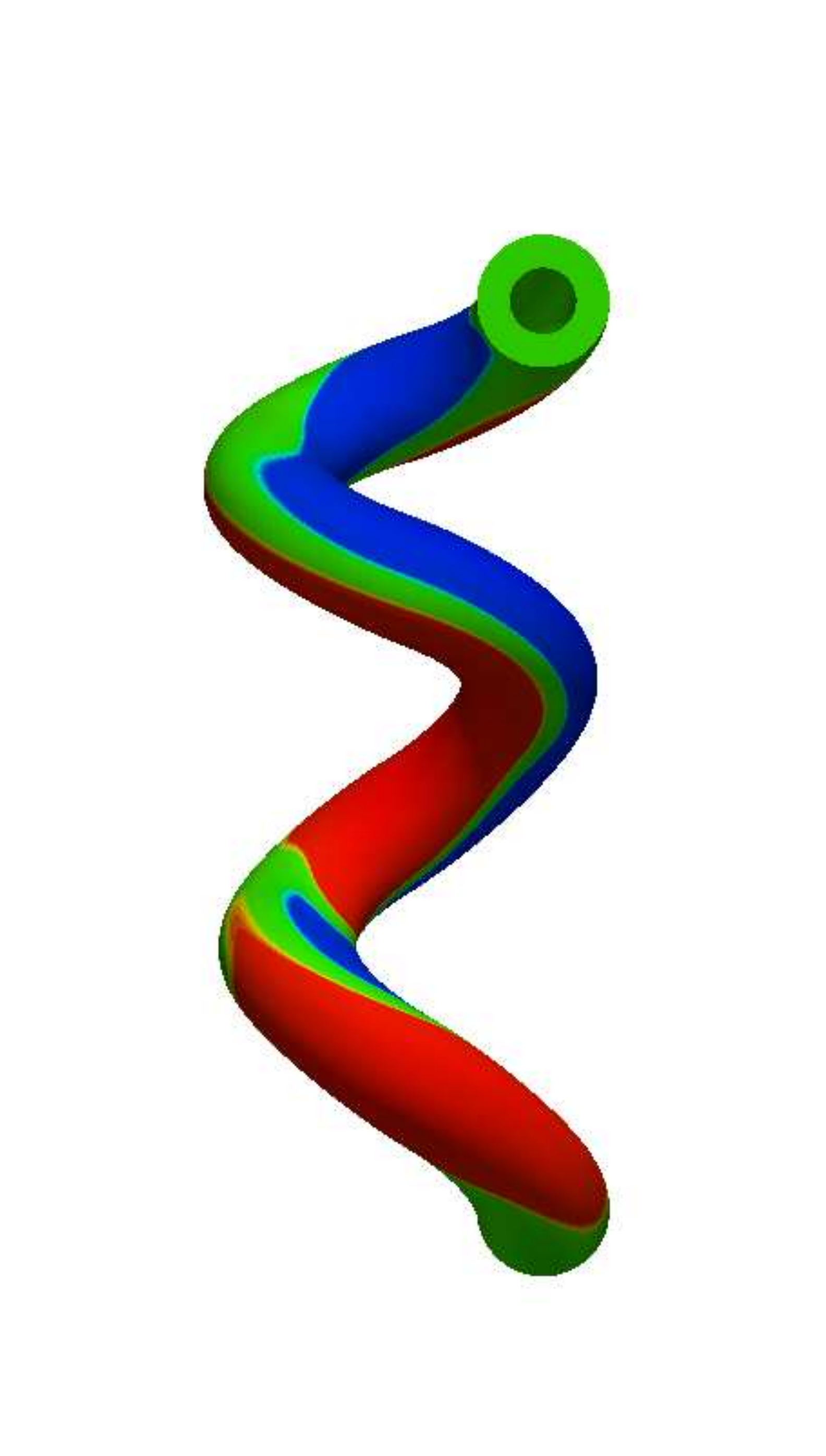}
\put(0,130){(e)}
\end{overpic}
}
\subfigure
{
\begin{overpic}[trim=35mm 5mm 35mm 35mm,clip, width=0.12\linewidth]  {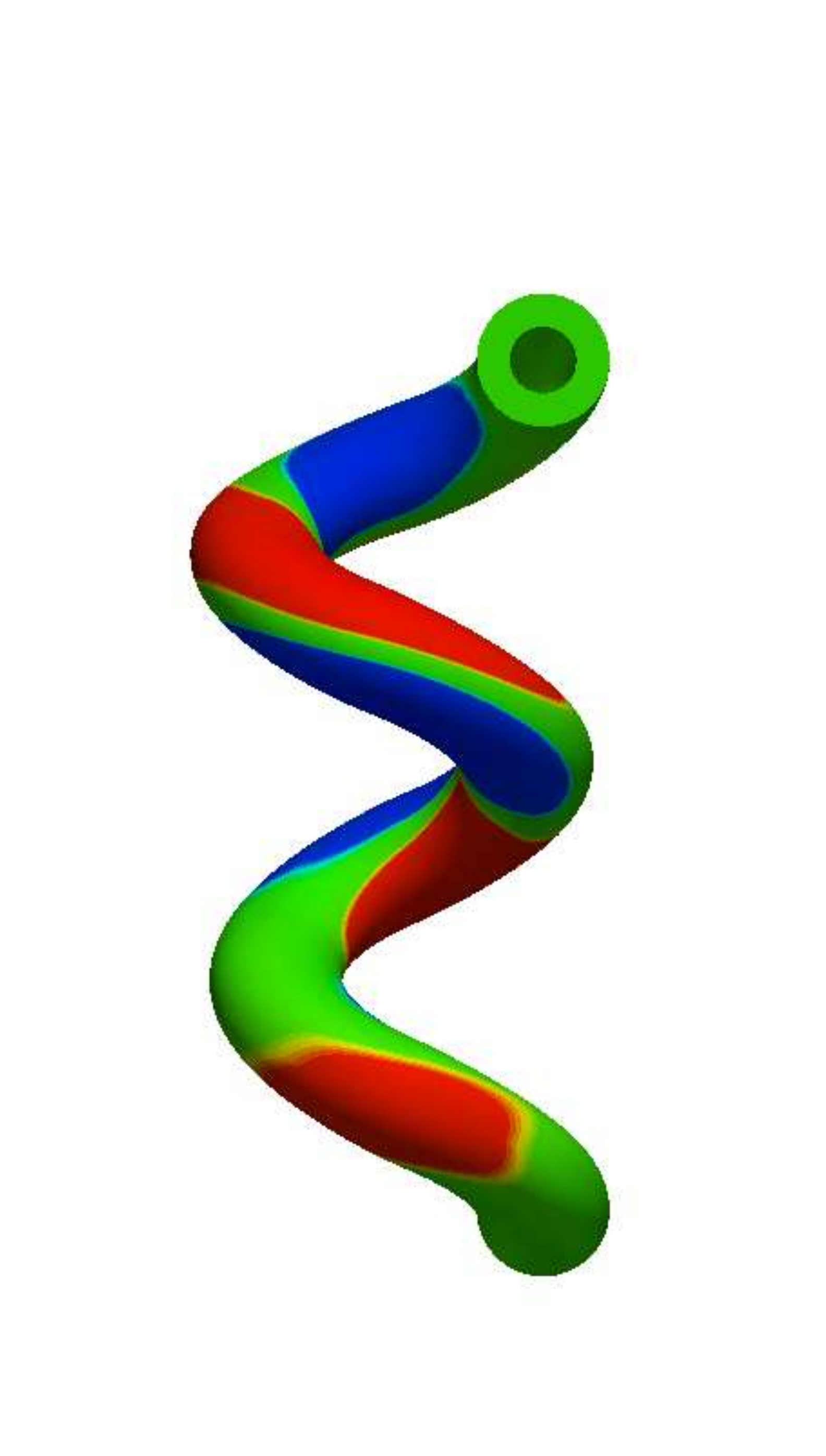}
\put(0,130){(f)}
\end{overpic}
}\\
\subfigure
{
\begin{overpic}[trim=40mm 5mm 25mm 35mm,clip, width=0.13\linewidth] {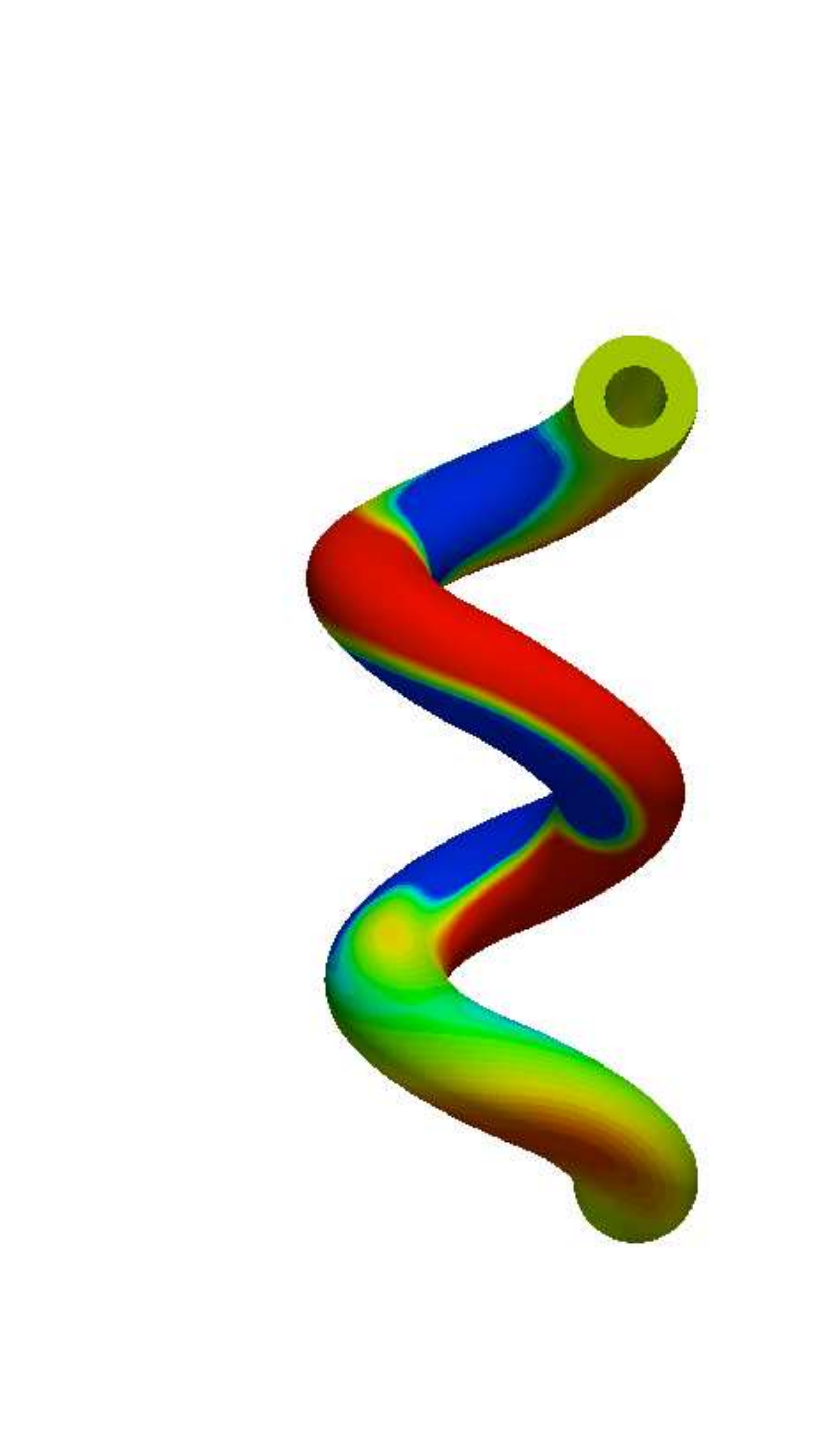}
\put(0,130){(a$^\prime$)}
\end{overpic}
}
\subfigure
{
\begin{overpic}[trim=40mm 5mm 25mm 35mm,clip, width=0.13\linewidth] {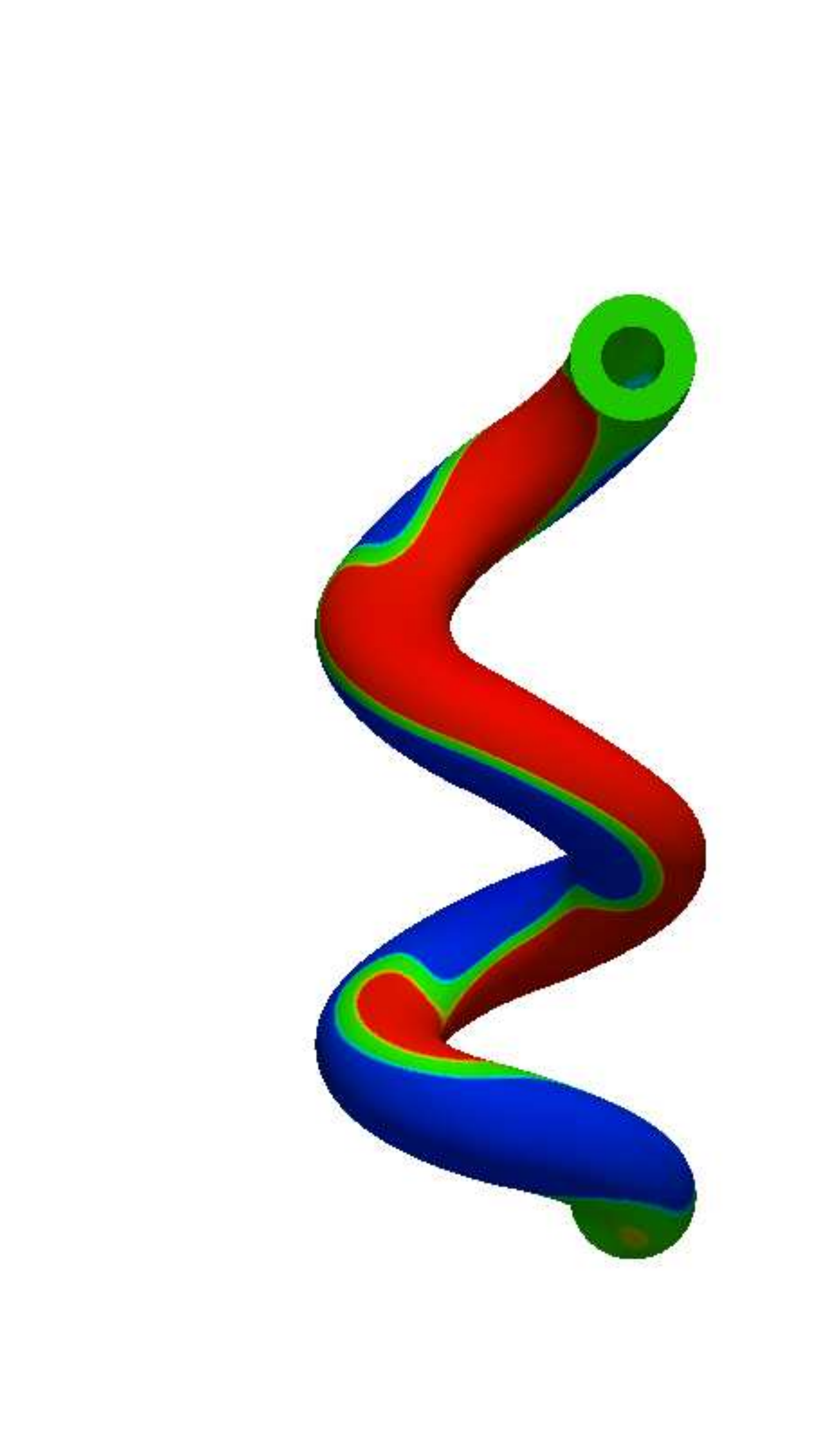}
\put(0,130){(b$^\prime$)}
\end{overpic}
}
\subfigure
{
\begin{overpic}[trim=40mm 5mm 25mm 35mm,clip, width=0.13\linewidth] {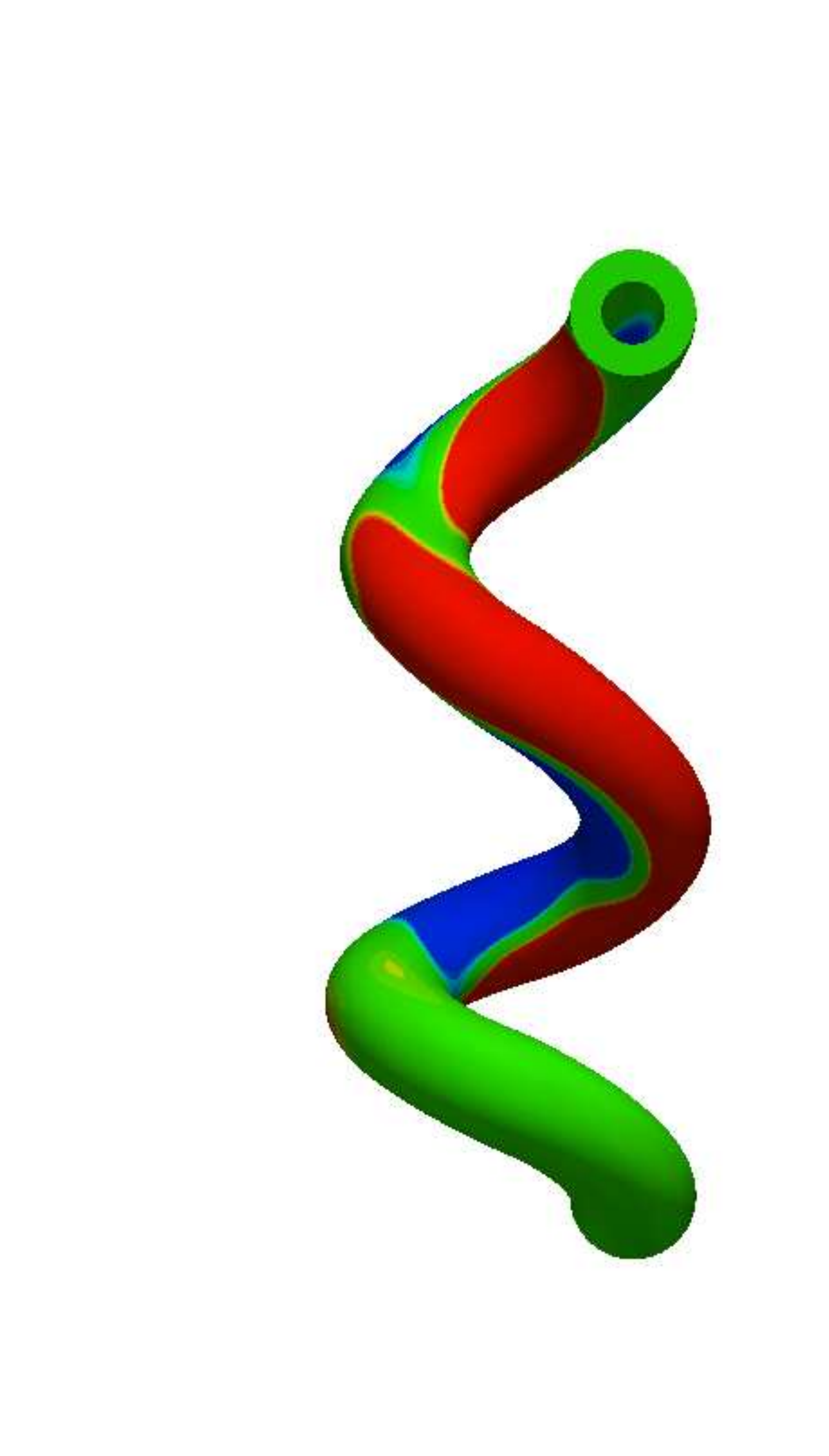}
\put(0,130){(c$^\prime$)}
\end{overpic}
}
\subfigure
{
\begin{overpic}[trim=40mm 5mm 25mm 35mm,clip, width=0.13\linewidth] {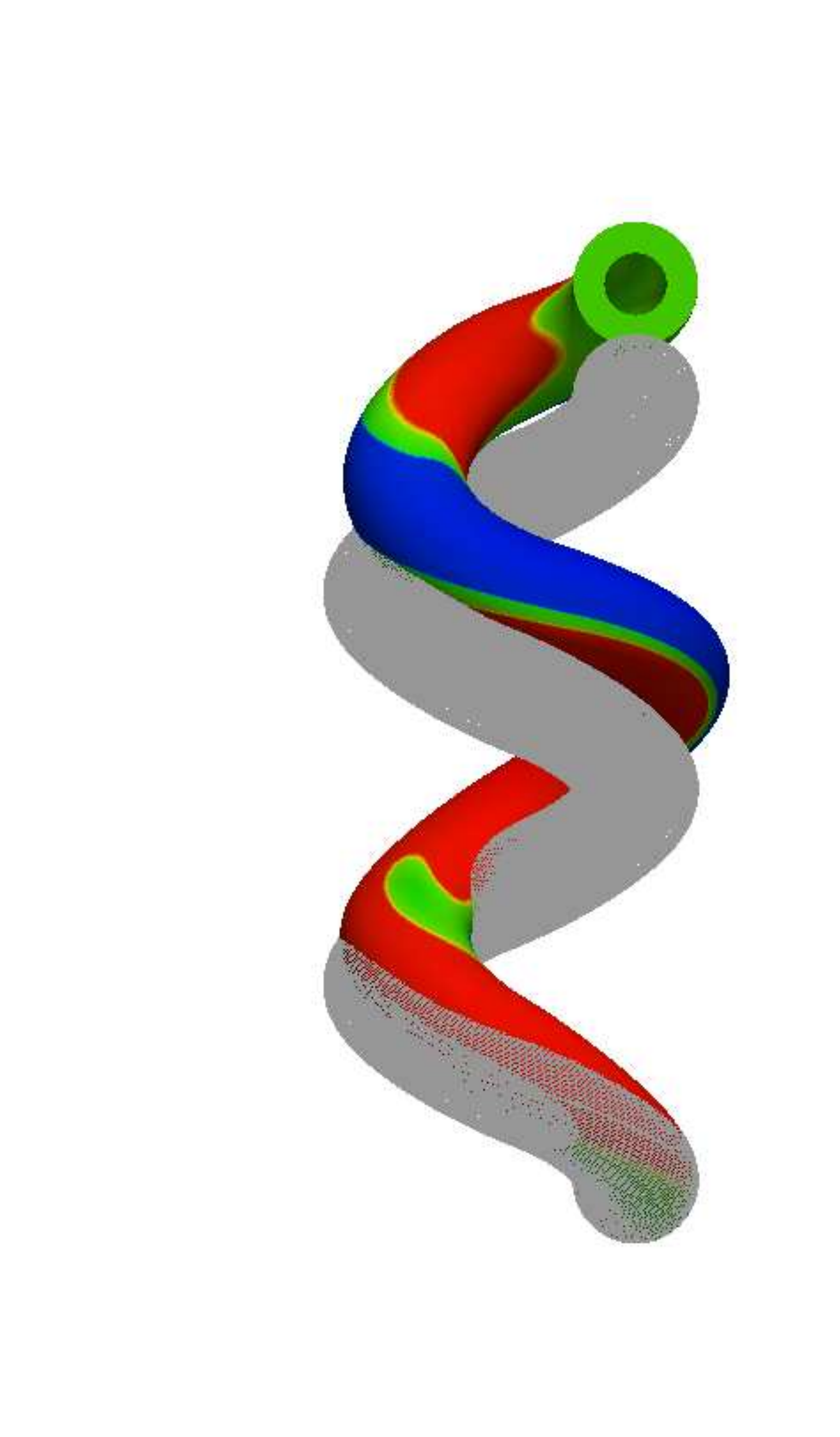}
\put(0,130){(d$^\prime$)}
\end{overpic}
}
\subfigure
{
\begin{overpic}[trim=40mm 5mm 25mm 35mm,clip, width=0.13\linewidth] {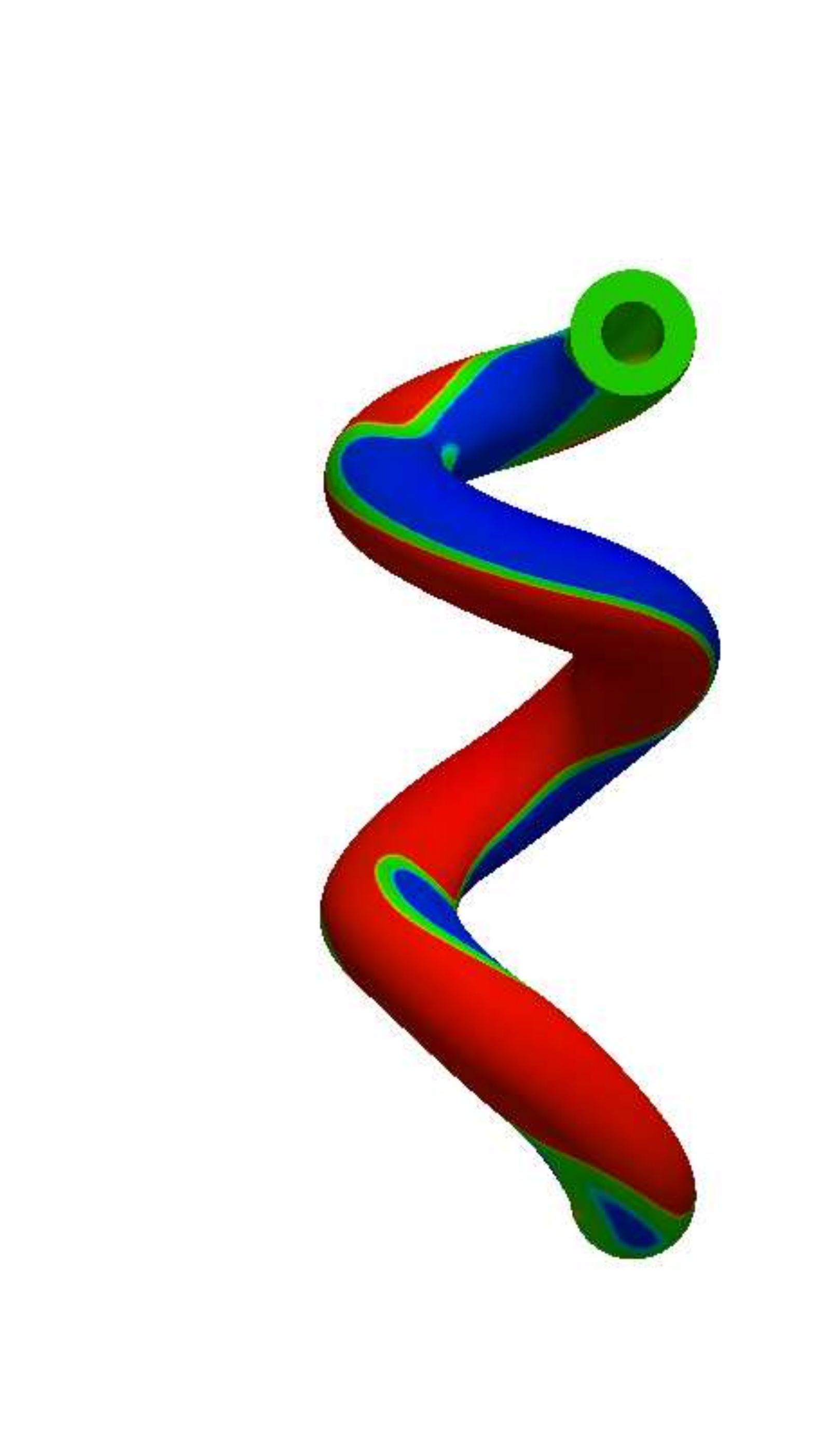}
\put(0,130){(e$^\prime$)}
\end{overpic}
}
\subfigure
{
\begin{overpic}[trim=40mm 5mm 25mm 35mm,clip, width=0.13\linewidth] {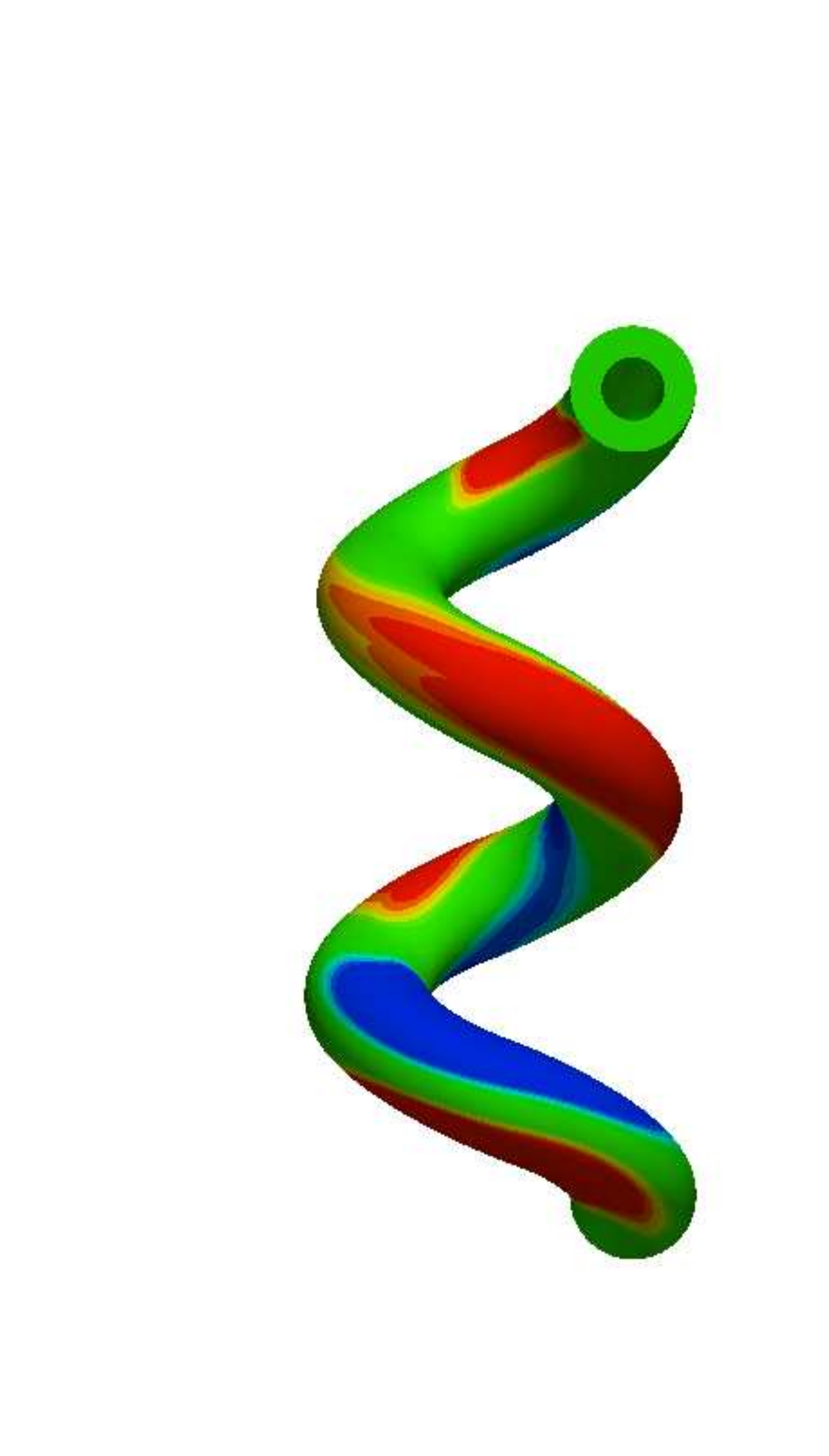}
\put(0,130){(f$^\prime$)}
\end{overpic}
}
\caption{(Color online) Microstructure evolution during pseudoelastic loading of a tubular nanostructured spring for two ramp loading-unloading cycles: (a)--(f) show the first cycle, (a$^\prime$)--(f$^\prime$) the second cycle (yellow, red, blue, and green colors represent austenite, M$_1$, M$_2$, and M$_3$ variants, respectively). The grey color represent undeformed mesh at the end of loadings in (d) and (d$^\prime$). }
\label{fig:SpringLoadingUnloading}
\end{figure}

Here we report the simulations results on a complex nanostructured SMA tubular spring specimen using the phase-field model. To the best of our knowledge, such results are obtained for the first time here for the case of tubular spring geometry. A two loop spring has the pitch of 190 nm and mean diameter of 60 nm. The outer and inner diameters of the tube are 82.5 nm and 41.25 nm. The tubular spring is meshed with 134, 16, and 266 \cone-quadratic NURBS basis functions in circumferential, radial, and helix directions, respectively. The top end of the SMA spring specimen is elastically loaded in the vertical direction and the bottom end is constrained with $\pmb{u}$ = $\pmb{0}$. The spring is consecutively loaded for two ramp loading-unloading cycles, each with 3 \% displacement with strain rate equivalent to 3$\times10^7/$s. The microstructures during the first and second cycles, at the same deformation (vertical displacement), are presented in Figs. \ref{fig:SpringLoadingUnloading}(a-f) and \ref{fig:SpringLoadingUnloading}(a$^\prime$-f$^\prime$), respectively.  The microstructure morphology is different at the same deformation of the spring during the two consecutive cycles. The study demonstrates the influence of dynamic loading on a microstructure evolution in consecutive loading-unloading cycles. 

The developed model can be  modified in a straightforward manner to incorporate the size dependent properties in the nanoscale specimens. Such accurate models can enhance our understanding of the MTs and thermo-mechanical behaviors of SMA nanostructures and provide guidelines for better device and application development.

\section{Conclusions} 
A new fully coupled thermo-mechanical 3D PF model has been developed that captures the underlying response of SMA nanostructures. It also qualitatively captures the important features of mechanical and thermal hystereses in pseudoelastic and shape memory effects during stress-induced transformations. The temperature variation during loading and unloading of SMA nanowire, due to  exothermic and endothermic processes, has been successfully captured as well. The local  temperature distribution acts as a signature of formation or movement of habit plane or domain wall. The model provides important information for better understanding of the MT mechanisms under dynamic loading conditions in SMA nanowires. 

\begingroup
\setstretch{0.8}
\bibliographystyle{unsrt}
\bibliography{IGA3DLettersarXiv2}
\endgroup

\end{document}